\documentclass[a4paper,fleqn,usenatbib]{mnras}

\usepackage{newtxtext,newtxmath}

\usepackage[T1]{fontenc}
\usepackage{ae,aecompl}

\usepackage{graphicx}	
\usepackage{amsmath}	
\usepackage{amssymb}	
\usepackage{bm}
\usepackage{longtable}

\usepackage[normalem]{ulem}

\title[Convection with misaligned gravity and rotation]{Convection with Misaligned Gravity and Rotation: Simulations and Rotating Mixing Length Theory}

\author[L. K. Currie et al.]{
Laura K. Currie,$^{1}$\thanks{E-mail: L.K.Currie@exeter.ac.uk}
Adrian J. Barker,$^{2}$
Yoram Lithwick$^{3}$
and Matthew K. Browning$^{4}$
\\
$^{1}$Department of Mathematics and Computer Science, College of Engineering, Mathematics and Physical Sciences, University of Exeter,\\
 EX4 4QF, UK\\
$^{2}$Department of Applied Mathematics, University of Leeds, LS2 9JT, UK\\
$^{3}$Center for Interdisciplinary Exploration and Research in Astrophysics (CIERA) \& Department of Physics and Astronomy,\\ Northwestern University, 2145 Sheridan Road, Evanston, IL 60208, USA\\
$^{4}$Department of Physics \& Astronomy, Stocker Road, University of Exeter, EX4 4QL, UK
}

\date{Accepted XXX. Received YYY; in original form ZZZ}

\pubyear{2019}

\begin{document}
\label{firstpage}
\pagerange{\pageref{firstpage}--\pageref{lastpage}}
\maketitle

\begin{abstract}
We present numerical simulations, using two complementary setups, of rotating Boussinesq thermal convection in a three-dimensional Cartesian geometry with misaligned gravity and rotation vectors. This model represents a small region at a non-polar latitude in the convection zone of a star or planet. We investigate the effects of rotation on the bulk properties of convection at different latitudes, focusing on determining the relation between the heat flux and temperature gradient. We show that our results may be interpreted using rotating mixing length theory (RMLT). The simplest version of RMLT (due to Stevenson) considers the single mode that transports the most heat. This works reasonably well in explaining our results, but there is a systematic departure from these predictions (up to approximately $30\%$ in the temperature gradient) at mid-latitudes. We develop a more detailed treatment of RMLT that includes the transport afforded by multiple modes, and we show that this accounts for most of the systematic differences. We also show that convectively-generated zonal flows and meridional circulations are produced in our simulations, and that their properties depend strongly on the dimensions of the box. These flows also affect the heat transport, contributing to departures from RMLT at some latitudes. However, we find the theoretical predictions of the multi-mode theory for the mid-layer temperature gradient, the root-mean-square (RMS) vertical velocity, the RMS temperature fluctuation, and the spatial spectrum of the heat transport at different latitudes, are all in reasonably good agreement with our numerical results when zonal flows are small.
\end{abstract}




\section{Introduction}
All stars, and many planets, are convective in parts of their interiors during some phases of their evolution (e.g.~\citealt{Kip2012}).  The heat transport associated with this convection, which must be calculated 
in order to construct a consistent stellar or planetary structure model, may be influenced by rotation or magnetism (e.g.~\citealt{Chandra1961}). 
But a quantitative understanding of the effects these have on the convection, and so ultimately on stellar/planetary evolution, is still lacking (e.g.~\citealt{Chabrieretal2007}).

A central challenge is that the convection typically involves turbulent motions occurring over a wide range of spatial and temporal scales, so that direct simulation of the underlying equations (namely those of magnetohydrodynamics, together with some form of energy equation and an equation of state) is not possible over evolutionary timescales (see, e.g., review in \citealt{KupMuth2017}).  Faced with this difficulty, one approach is to simulate the problem for a comparatively brief interval in a star or planet's life, aiming to capture as much of the dynamics as possible with finite computational resources; another is to parameterise the convection in a way that can be computed in the course of an evolutionary calculation.  In the former case, choices must still be made about which scales to resolve: for example, some models encompass a full spherical domain, explicitly following the evolution of the largest-scale flows and relying on sub-grid-scale descriptions of smaller-scale flows; others model a more limited portion of the interior, but can then afford to resolve turbulent flows at smaller scales.  Whether on global or local scales, such simulations provide some insight into the complex dynamics that can occur when rotation or magnetism influence the convection.  Meanwhile, the parameterised models, of which "mixing length theory'' is the most famous and widely-used example (e.g.~\citealt{BV58,GoughWeiss1976,Kip2012}), aim principally to provide a formula for the temperature gradient necessary for convection to carry a certain flux (or vice versa), which can then readily be included in a structural or evolutionary model.  These models do not, in their usual formulation, include the effects of rotation or magnetism at all (though some exceptions to this rule are noted below).  

In this paper, we explore the effects of rotation on convection using both of these broad approaches.  We conduct a large survey of 3D simulations in localised, Cartesian domains tilted at some angle with respect to the rotation vector, extending prior work on this subject in a manner described below.  These box simulations serve as idealised representations of a small part of a rotating star or planet, situated at various latitudes. We use these to analyse the rich variety of phenomena that occur as the rotation rate and latitude are changed, and to compute how the temperature gradient established by the convection varies with these parameters. We compare the results of these calculations to expectations from semi-analytical theory: in particular, we argue that a multi-mode theory developed here, based on the "rotating mixing-length theory'' of \cite{Stevenson1979} (hereafter S79) provides a reasonably good description of the dynamics at most latitudes and rotation rates.

We begin here by briefly outlining some of the prior simulations and theory that motivate and guide our work.

\subsection{Prior studies of rotating convection}

Arguably the most realistic approach to modeling a rotating stellar/planetary convection zone is to solve the underlying equations numerically within a spherical computational domain.  Simulations of this type naturally capture the global geometry and the largest-scale flows, and have been used for decades (e.g.~\citealt{Gilman1975}) to assess how convection is influenced by rotation, and how this in turn affects the driving of zonal flows and the establishment of magnetic fields.  The increase of computing power  in recent times has led to substantial progress with such models, with the latest models (see, e.g.,~\citealt{Gastine2016,Hotta2016,Kapyla2017,Strugarek2018}) resolving increasingly turbulent flows over a broad range of spatial and temporal scales. (See, for example, review in \citealt{BrunBrowning2017}.)   Such models reveal that the convective transport of heat and angular momentum is a function of rotation rate and of latitude (e.g.,~\citealt{Raynaudetal2018}), in broad accord with some of the theoretical expectations described below.  However, these simulations are still very computationally demanding, rendering it difficult to probe large regions of parameter space systematically.  

Given the finite computational resources available, often a local model is used instead: here, a small region of a spherical body is modelled, so that the body's curvature can be neglected and Cartesian coordinates employed (see, e.g., discussions in \citealt{Vallis2006}, and review of local convection calculations in \citealt{Nordlund2009}). The majority of local models take the rotation and gravity vectors to be aligned (e.g.,~\citealt{Cattaneo1991,Barkeretal2014}, hereafter Paper I), but as an intermediate step between local aligned models and global models, the so-called tilted f-plane can be considered in which gravity and rotation are taken to be misaligned. 

Less attention has been given in the literature to rotating convection with tilted rotation vectors than in the aligned case, but there are still some notable studies. The linear theory was analysed by \citet{H79}, \citet{Stevenson1979} and \citet{FG78}, in the absence of viscosity and thermal diffusion, and \cite{H80} including diffusive effects. 
\citet{JK98} performed an asymptotic analysis for nonlinear rapidly rotating convection and derived predictions that agreed well with the laminar simulations of \citet{HS1983}. Those simulations adopted the Boussinesq approximation (e.g.,~\citealt{SV1961}), which neglects density fluctuations except where multiplied by gravity; this can be regarded as assuming a layer depth that is small compared to the scale height, and slow flows compared with the sound speed. Later studies have also simulated anelastic \citep{CurrieTobias2016} and fully compressible convection with tilted rotation vectors
to explore the effect of rotation on mean flows, convective transport, and convective overshooting at mid-latitudes in the Sun \cite[see e.g.,][]{Brummell1996,Brummell1998,Brummell2002,Kapyla2004,Chan2007}. 

Rotating convection, whether occurring in localised domains or in global ones, is known to generate mean flows. For example, \citet{HS1983} demonstrated the existence of self-consistent mean flows in tilted models at moderate rotation rates.
At the poles, convection has been observed to generate large-scale vortices or horizontal jets (e.g. \citealt{Chan2007,Kapyla2011,Guervilly2014,Favier2014,Rubio2014,Guervilly2017,Julien2017}), depending on the horizontal aspect ratio of the simulated domain. The occurrence of these flows in simulations of rotating convection with misaligned gravity and rotation, and their resulting effects on the heat transport, remain to be explored in detail. However, their presence in compressible convection was noted by \citet{Chan2007} and \citet{Mantereetal2011} and more recently, some progress has been made in the Boussinesq case by \cite{Novi2019}.

\subsection{Heat transport and prior theory}

From the point of view of stellar or planetary structure, the primary purpose of a convection theory is to provide an estimate of the temperature (or entropy) gradient needed to carry a given flux. This provides partial motivation for many prior studies that have sought to constrain the temperature gradient established by turbulent convection, with and without rotation, either by simulation or through analytical theory. 
In the fluid dynamics literature, this is typically expressed as a relation between the Nusselt number, $Nu$ (a dimensionless measure of the flux carried by convection, relative to the conductive flux) and the Rayleigh number, $Ra$ (quantifying buoyancy driving relative to dissipation). 
These quantities can be defined either globally (e.g., in terms of the total temperature or entropy contrast across a layer) or locally (e.g., by reference to the local temperature or entropy gradient), depending on the setup under consideration.  

For example, non-rotating convection is often argued to approach the diffusion-free relation $Nu \propto (Ra Pr)^{1/2}$ for very large $Ra$ \citep{Kraichnan1962,Spiegel1971}, here employing typical definitions for $Nu$ and $Ra$ appropriate for fixed-temperature boundary conditions that depend on the temperature difference between the two boundaries, and where $Pr=\nu/\kappa$ is the Prandtl number (ratio of kinematic viscosity $\nu$ to thermal diffusivity $\kappa$). 
Analysis of the heat transport is made complicated by the fact that in experiments or numerical simulations, much of the temperature drop typically occurs in thin boundary layers near the top and bottom of the convective region, within which heat is transported primarily by conduction. 
Indeed, the experimentally observed relationship between temperature drop and heat flux can be approximately accounted for by considering only the behavior of the boundary layers (e.g.~\citealt{Malkus1954}) which leads to $Nu\propto Ra^{1/3}$. 
In this non-rotating regime, we might regard the convective transport through the domain as being "throttled" by the boundary layers.  However, when rotation is present
simulations have suggested that the $Nu(Ra)$ scaling is steeper \citep[see e.g.,][]{Kingetal2009, KingStellmachBuffett2013}, but as the buoyancy driving is increased the simulations latch onto a diffusion-free scaling (e.g.,~\citealt{Tilgner09,Stellmach2014}) before losing their rotational influence at large $Ra$.
Other progress has been made by modeling reduced sets of equations, valid in an appropriate asymptotic regime (e.g., corresponding to rapid rotation); see for example \citet{JK98}.

In stellar astrophysics, the transport by convection is typically parameterised using mixing-length theory (MLT); \citep[see, e.g.,][]{BV58, GoughWeiss1976}.  Broadly, the theory models convection as parcels of fluid that travel a specified length (the mixing length) before giving up heat to their surroundings.
MLT is fundamentally a local theory, relating the value of the superadiabatic temperature gradient at a specific point to the flux at that point; its predictions for these quantities are diffusion-free, and can be cast in the form $Nu \propto (RaPr)^{1/2}$ given suitable local definitions of $Nu$ and $Ra$ \citep[see, e.g.,][]{GoughWeiss1976}.

Despite its simplicity, MLT has been remarkably successful in modelling the gross structures of stars and gaseous planets \citep{Baraffeetal2015}. However, standard MLT suffers from many well-known limitations.
In its usual formulation, it omits many important effects altogether, including rotation and magnetic fields. The former is unlikely to be important in modifying the structure of a star \citep{IrelandBrowning2018}, but may impact mixing and the generation of differential rotation, for example. Standard MLT is also too simplistic to model overshooting and time-variability
accurately \citep[e.g.,][]{Renzini1987,Arnettetal2019}, or other effects such as the
asymmetries between upflows and downflows \citep{Nordlund2009}.
Several authors have considered other approaches for modeling stellar convection.  These include, for example, the "full spectrum turbulence" model of \citet{Canuto1996}, or the Reynolds stress models of \citet{Xiong1978, Xiong1989}, \citet{Xiongetal1997} and \citet{Canuto2011}.  Such models provide a more physically-motivated description of convection, including its transport of tracer particles, of heat, and of momentum, and have been used (for example) to study overshooting in massive stars.

 As an intermediate step between (for example) closure-based models and classical MLT, it is possible to construct convection theories that largely share the simplicity of MLT, but seek to incorporate some of the physical effects missing from it in a physically consistent way.  As an example of this approach, S79 (see also \citealt{FG78}) proposed a physically-motivated extension of mixing length theory that incorporates rotation, which we will hereafter refer to as rotating mixing length theory (RMLT). This is a simple theory for rotating convection, based on the idea that turbulent convection is dominated by modes that are well described by linear theory (without viscosity and thermal diffusion), but with amplitudes determined by the (nonlinear) amplitude limiting criterion that each mode saturates when its growth rate balances its nonlinear cascade rate. S79 derived the simple analytical predictions of this theory when a single mode -- the one which maximises the heat flux -- dominates the heat transport. (A similar saturation prescription for stellar convection has been adopted by \cite{Lesaffre2013} and \cite{Jermyn2018} for implementation in stellar evolution codes, which gives different predictions in its current formulation.)

S79's RMLT is derived in more detail and extended in Section \ref{MLT} below; from the point of view of stellar structure its most significant prediction is that the temperature gradient in the bulk of the convection zone, in the limit of rapid rotation and considering a region near the poles, scales with the rotation rate to the four-fifths power. That is, more rapid rotation constrains the motions and requires a higher temperature gradient (less efficient convection) in order to carry the same heat flux.  The theory also provides formulae for the vertical velocity and temperature fluctuations, and for the horizontal wavenumber of the modes that dominate the heat transport. A different line of reasoning based on \cite{Kraichnan1962}, followed by \cite{Julien2012}, yields the same basic scaling of temperature gradient with rotation rate, as follows: 
suppose $Nu \propto (Ra/Ra_c)^\alpha Pr^\beta$ for some $\alpha$ and $\beta$ to be determined, together with our knowledge that the critical Rayleigh number $Ra_c\propto Ek^{-4/3}$ from linear theory. ($Ek$ is the Ekman number, defined in (\ref{tradnondim}), and is a measure of the viscous forces relative to the Coriolis force). Then, if we assume that the heat flux is independent of the diffusivities, it is possible to obtain $\alpha=3/2$ and $\beta=-1/5$ \citep{Julien2012}. 
The resulting scaling for $Nu$ was observed in simulations using an asymptotically reduced model for rapidly rotating Rayleigh-B\'{e}nard convection at the poles, in which the bulk and not the boundary layers dominate the transport, and this was later confirmed by the DNS of \cite{Stellmach2014}. In terms of the flux Rayleigh number $Ra_F = Ra Nu$, this yields a turbulent heat transport scaling\footnote{Note that this scaling also in principle applies with misaligned gravity and rotation if $Ek$ is defined as in Eq.~\ref{tradnondim}.} of 
$Nu \propto Pr^{-1/5} Ra_f^{3/5} Ek^{4/5}$. Despite the entirely different derivation, this scaling is equivalent to the single-mode prediction for $dT/dz$ from RMLT. These predictions are also similar to the inertia-free scalings extensively discussed in the planetary sciences and geodynamo communities \citep[e.g., see][]{IngPoll1982,Aub01,GilletJones2006,Guervilly2019}.

The scaling predictions of RMLT and related theories have found support from prior simulations, both in spherical shells and in localised domains.  For example, in recent simulations of rotating Boussinesq convection in a spherical geometry, \citet{Gastine2016} suggested that the diffusion-free predictions of RMLT matched the numerical data for sufficiently rapid rotation and small diffusivities. In Cartesian domains, Paper I used Boussinesq direct numerical simulations (DNS) of rotating convection to verify the scaling predictions of RMLT over several orders of magnitude in rotation rate. That paper argued that the bulk convective properties (i.e., away from the boundary layers) were indeed approximately independent of the diffusivities, insofar as this could be probed with the simulations. However, in these simulations the rotation vector $\bm\Omega$ was aligned with gravity, which restricts the application of this model to the polar regions of a star or planet. The applicability of RMLT to non-polar latitudes in a star or planet is not yet known.
More recently, \citet{Andersetal2019} found that a different $Nu$--$Ra$ scaling better described their simulations of compressible convection in a rotating, Cartesian domain. It is not entirely clear why their simulations appear to exhibit different heat flux scalings than in some previous work (e.g., Paper I), but one possibility is that the definition of $Nu$ in \citet{Andersetal2019} includes boundary layers that may dominate the scaling behaviour at high $Ra$, whereas (for example) Paper I focussed on measuring properties in the middle regions of the simulated convection zone (i.e., neglecting boundary layers).

In this paper we are interested in the effects of rotation and large scale flows on convection, and in particular the temperature gradient at different latitudes (so that gravity and rotation are misaligned). Furthermore, we are concerned with whether the results of this study can be understood within the framework of RMLT.
To this end we use two different numerical setups (described in section \ref{model}) and investigate the basic properties of the developed convection in simulations with gravity and rotation misaligned (section \ref{sims}). In both cases we employ the Boussinesq approximation: this is of course less realistic than modelling the fully compressible equations, but is more computationally tractable and allows us to isolate the effects of rotation from other influences (like the density stratification) that are not our primary focus.  By focusing on a localised Cartesian domain, we are able to access parameter regimes that are more difficult to reach in global-scale models, and to conduct limited surveys of the vast parameter space available.  We focus our discussions largely on the temperature gradient, and seek a physical understanding for its dependence on rotation rate and latitude via RMLT.
 We choose to compare to RMLT because it provides a compelling and simple theory that well describes rotating convection at the poles, and also because many of the effects considered in more complex models of convection (e.g., overshooting, or asymmetry between upflows and downflows) are not relevant for the simple setup considered here.
The RMLT arguments are presented in section {\ref{MLT}} and tested against the numerical simulations. The effect of zonal flows are discussed in section \ref{Meanflows} before a discussion of our results in section \ref{conc}.

\section{Model setup and equations}\label{model}
\begin{figure}
    \centering
    \includegraphics[scale=0.6]{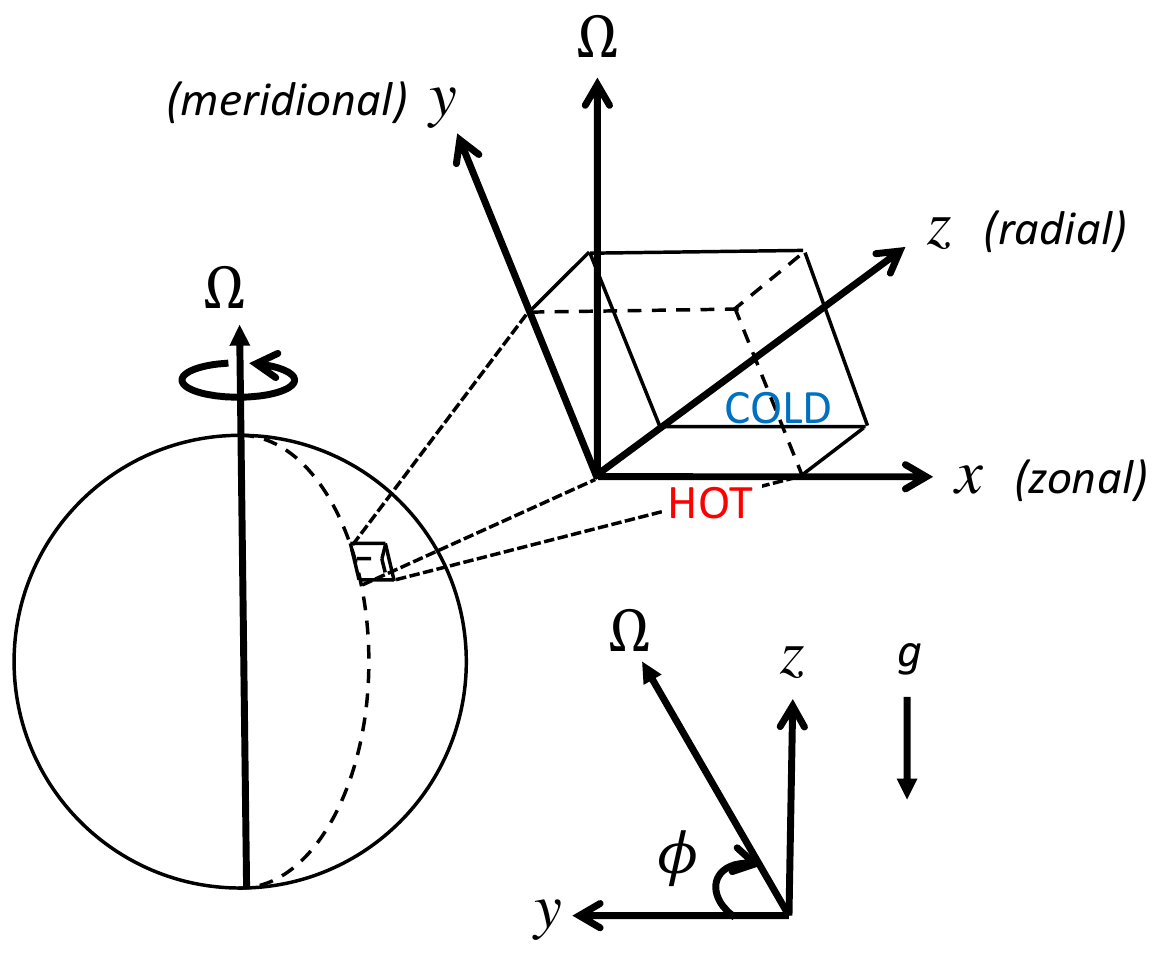}
    \caption{A schematic diagram showing the geometry of the model used in simulations. A small Cartesian box that is rotating at a rate $\Omega$ is taken at a latitude $\phi$. $x$ points eastwards, $y$ northwards and $z$ vertically upwards; the rotation axis is therefore taken to lie in the $y$-$z$ plane. An adverse temperature gradient in the $z$ direction drives convection.}
    \label{schematic}
\end{figure}

We consider two different and complementary numerical approaches to simulate rotating turbulent convection in a Cartesian domain with a rotation vector $\boldsymbol{\Omega}=\Omega(0,\cos\phi,\sin\phi)$ that is in general tilted with respect to $z$ (where $\Omega$ is the rotation rate and $\phi$ is the latitude, so that $\phi=90^{\circ}$ corresponds to the pole). 
We consider $x$ to point eastwards, $y$ to point northwards and $z$ to point upwards. The geometry of this setup is illustrated in Fig. \ref{schematic}.
The first approach (setup A) uses a Rayleigh-B\'enard-type setup where convection is driven by imposing the heat flux at one of the boundaries in $z$ instead of fixing the temperature. The second approach (setup B) drives the convection using heating/cooling layers adjacent to the boundaries in $z$, as previously used in Paper I. The advantage of using two different numerical codes that drive the convection differently
is that this allows us to determine whether the bulk convective properties are independent of the way convection is driven, and to ensure that our main results are robust.

In both cases, the governing equations are those of rotating Boussinesq thermal convection:
\begin{equation}\label{momeq}
\frac{\partial \bm u}{\partial t} + (\bm u\cdot \nabla)\bm u + 2\bm\Omega \times \bm u = - \nabla p + T\bm e_z + \nu \nabla ^2 \bm u,
\end{equation}
\begin{equation}
\frac{\partial T}{\partial t} + (\bm u\cdot \nabla)T= q +\kappa \nabla ^2 T,
\end{equation}
\begin{equation}\label{incompeq}
\nabla \cdot \bm u =0, 
\end{equation}
where $\bm{u}=(u_x,u_y,u_z)$ is the fluid velocity, $p$ is a pressure and $T$ is a scaled temperature which can be thought of as a buoyancy variable and has the units of an acceleration. More specifically, $T=g\alpha\Delta T_{real}/T_0$, where $g$ is the magnitude of gravitational acceleration (which points downwards), $\alpha$ is the coefficient of thermal expansion and $\Delta T_{real}/T_0$ is the fractional difference in the real temperature relative to a reference value, $T_0$. 
In setup B, internal heating/cooling is implemented by choosing $q(z)$, and further details are given in section \ref{setupB}.

Our computational domain is a Cartesian box with $x\in[0,L_x]$, $y\in[0,L_y]$, $z\in[0,L_z]$. In both setups $L_z$ is fixed, and takes a value that will be specified below for each of our setups. On the other hand, $L_x$ and $L_y$ are both varied to explore the effects of varying the horizontal domain sizes and aspect ratio on the convection and the resulting mean flows. We aim to choose $L_x$ and $L_y$ so that we can simulate multiple wavelengths of the dominant convective modes.

\subsection{Setup A: Rayleigh-B\'enard-type}
For these simulations, the flux is fixed through the domain via the temperature boundary conditions. In particular, we impose
\begin{eqnarray}\label{setupAbcs}
T=0 \;\;\;\;\text{at} \;\;\;\;z=0 \;\;\; \& \;\;\; \partial_z T=-F/\kappa \;\;\;\text{at}\;\;\; z=L_z,
\end{eqnarray}
for a given flux, $F$, and thermal diffusivity, $\kappa$. For the velocity boundary conditions, we assume impenetrable, free-slip boundaries so that 
\begin{eqnarray}
\partial_z u_x=\partial_z u_y=u_z=0 \;\;\;\; \text{at} \;\;\;\; z=0 \;\;\& \;\;L_z.
\end{eqnarray}
We use periodic boundary conditions in $x$ and $y$ for all variables.
\citet{HS1983} used a similar Rayleigh-B\'enard-type setup but they considered fixed temperature conditions at both boundaries and enforced no-slip boundary conditions on the velocity. Our simulations differ further from \citet{HS1983} in that we consider much faster rotation and lower diffusivities than the regime they probed in their study. 

The simulations of this setup were carried out using Dedalus (http://ascl.net/1603.015; http://dedalus-project.org; \citealt{Dedalus2019}), a pseudo-spectral code with implicit-explicit timestepping. Typically, a 2nd-order Crank-Nicholson, Adams-Bashforth scheme was used for the timestepping, with a CFL condition restricting the timestep. A Chebyshev spectral method was adopted in $z$ and a Fourier method in $x$ and $y$. Dealiasing was implemented using the 2/3 rule. Most simulations used 192 grid points in each direction, but for larger boxes more grid points were used (see Table~\ref{Table1} for details). We fix $L_z=1.2H$ for all simulations using setup A, where $H$ is the depth of the convection zone in setup B (see section \ref{setupB}). This choice was made to allow a more direct comparison between the two setups.

\subsection{Setup B: heating/cooling layers}\label{setupB}
Again, we adopt periodic boundary conditions in $x$ and $y$ for all variables, and stress-free, impenetrable conditions in $z$. The thermal boundary conditions in $z$ are zero temperature at the bottom ($z=0$) and insulating at the top ($z=L_z$). Specifically, we impose
\begin{eqnarray}
T=0 \;\;\;\;\text{at} \;\;\;\;z=0 \;\;\; \& \;\;\; \partial_z T=0 \;\;\;\text{at}\;\;\; z=L_z.
\end{eqnarray}
Convection is driven by imposing internal heating and cooling so that fluid is heated at the bottom of the box in a zone of depth $\Delta=0.2H$, and cooled by an equal amount at the top, with no heating/cooling in the middle "convection zone", which has depth $H$ so that $L_z=H+2\Delta=1.4H$. Following Paper I, we adopt 
\begin{eqnarray}
\label{hcprofile}
\nonumber
q(z) = \frac{F}{\Delta} \begin{cases}
 		  1+\cos \left(\frac{2\pi (z-\Delta /2)}{\Delta} \right) & \text{if } 0\leq z \leq \Delta , \\
   		 0  & \text{if } \Delta < z <L_{z}- \Delta , \\
		-1-\cos \left(\frac{2\pi (z-L_{z}+\Delta/2)}{\Delta } \right) &  \text{if } L_{z}-\Delta  \leq z \leq L_{z}, \\
  	 \end{cases}
\end{eqnarray}
with integrated heating of $F$ in the top and bottom layers. Together with our boundary conditions, this constrains the total heat flux $F$ in the convection zone in a steady state.

We use the efficiently-parallelised spectral element code Nek5000 \citep{Nek5000}, which was previously used in Paper I. This method partitions the domain into a set of $\mathcal{E}$ non-overlapping elements, and within each element the velocity components and the pressure are represented as tensor product Legendre polynomials of order $\mathcal{N}$ and $\mathcal{N}-2$, respectively, defined at the Gauss-Lobatto-Legendre and Gauss-Legendre points. Such a method has algebraic convergence with increasing $\mathcal{E}$, and spectral (exponential) convergence with increasing $\mathcal{N}$ (for smooth solutions), with the total number of grid points in 3D being $\mathcal{EN}^3$. Since the grid points are non-uniformly spaced, whenever we wish to compute Fourier spectra, we first interpolate the numerical data to a uniform grid using in-built routines in Nek5000.

Temporal discretisation is based on a semi-implicit formulation, where the nonlinear and Coriolis terms are treated explicitly, and the viscous and pressure terms implicitly. In particular, we use a 2nd-order characteristics-based timestepper for the explicit terms and a 2nd-order backward-difference formula for the viscous and pressure terms, with a variable time-step determined by a target CFL number. The nonlinear terms are fully de-aliased by using a polynomial order that is 3/2 larger for their evaluation. Our typical resolution is $\mathcal{E}=20^3$ and $\mathcal{N}=10$ (15 for the nonlinear terms), unless otherwise specified in Table~\ref{Table1}, where we list the parameters for all of our simulations.

\subsection{Non-dimensionalisation}
For simulations using either setup we fix the total heat flux through the domain. In setup A, this is achieved through the boundary conditions (\ref{setupAbcs}) and in setup B this is achieved by the heating/cooling function $q(z)$ in addition to the boundary conditions. The total heat flux is given by
\begin{equation}
    F=-\kappa\frac{{\rm d}\langle T\rangle_{xy}}{{\rm d}z} + \langle u_zT \rangle_{xy}
\end{equation}
where the angled brackets with subscripts represent averaging over the directions indicated (i.e., horizontal, here). In a steady state, the time-averaged $F$ is independent of height in the convection zone.

Following Paper I we set $F=1$ and $H=1$.
In other words, we measure lengths in units of $H$ and times in units of $H^{2/3}/F^{1/3}$. This choice was made because it allows us to compare with the predictions of RMLT most simply.

For each simulation we also specify $\Omega$, $\phi$, $\nu$ and $\kappa$ (along with $L_x$ and $L_y$). These quantities are related to the more traditional dimensionless numbers in the following way
\begin{eqnarray}\label{tradnondim}
Ra_F=\frac{FH^4}{\kappa^2\nu}, \;\;\;\;  Ek=\frac{\nu}{2\Omega\sin\phi H^2}, \;\;\;\; Pr=\frac{\nu}{\kappa},
\end{eqnarray}
where $Ra_F$ is a flux based Rayleigh number, $Ek$ is the Ekman number and $Pr$ is the Prandtl number.
In terms of these dimensionless numbers, the dimensionless time, $t$, in our code can be expressed as $t=t_{visc}(Ra_F/Pr^2)^{1/3}$ where $t_{visc}$ is the time expressed in units of a viscous timescale ($H^2/\nu$).

To keep our voyage through parameter space manageable, we set $\nu=\kappa$ throughout, i.e.~$Pr=1$, deferring a study of the possible dependence of the bulk properties (or mean flows) on $Pr$ to future work. 
We also fix $\nu$ for most of the simulations with a given $\Omega$, though we do vary $\nu$ separately for a subset of simulations. 
We perform simulations for a range of $\Omega$ and vary $\phi$ (in addition to $L_x$ and $L_y$). 
Most of the simulations are in a regime where they are strongly influenced by rotation, i.e., they have Rossby numbers (defined by $Ro=\frac{u_{z,rms}}{2\Omega H}$) much less than one. As an example, consider case 10A45b: as listed in Table~\ref{Table1}, this has $\Omega=10$ and $u_{z,rms}=0.56$, so that $Ro=0.03$.  Alternative measures of the rotational influence, e.g., the "convective Rossby number" $Ro_c = Ek(Ra/Pr)^{1/2}$, also typically yield values less than unity -- e.g., for the same case, $Ro_c$ is approximately 0.26, after calculating the traditional Rayleigh number $Ra$ via $Ra=Ra_F/Nu$ and $Nu=F/\kappa|dT/dz|$ (where the temperature gradient is taken over the middle one third of the domain) -- likewise indicating that rotation plays a significant role. We focus on this regime because it allows us to directly test the predictions of RMLT.
The values of parameters used in our simulations are given in Table~\ref{Table1} from which corresponding $Ek$ and $Ra_F$ can easily be computed.
In general, the supercriticality changes between different simulations -- we comment on this further in section \ref{supercrit}.

\section{Tilted convection in the nonlinear regime}\label{sims}
\begin{figure}
    \centering
    \includegraphics[scale=0.6]{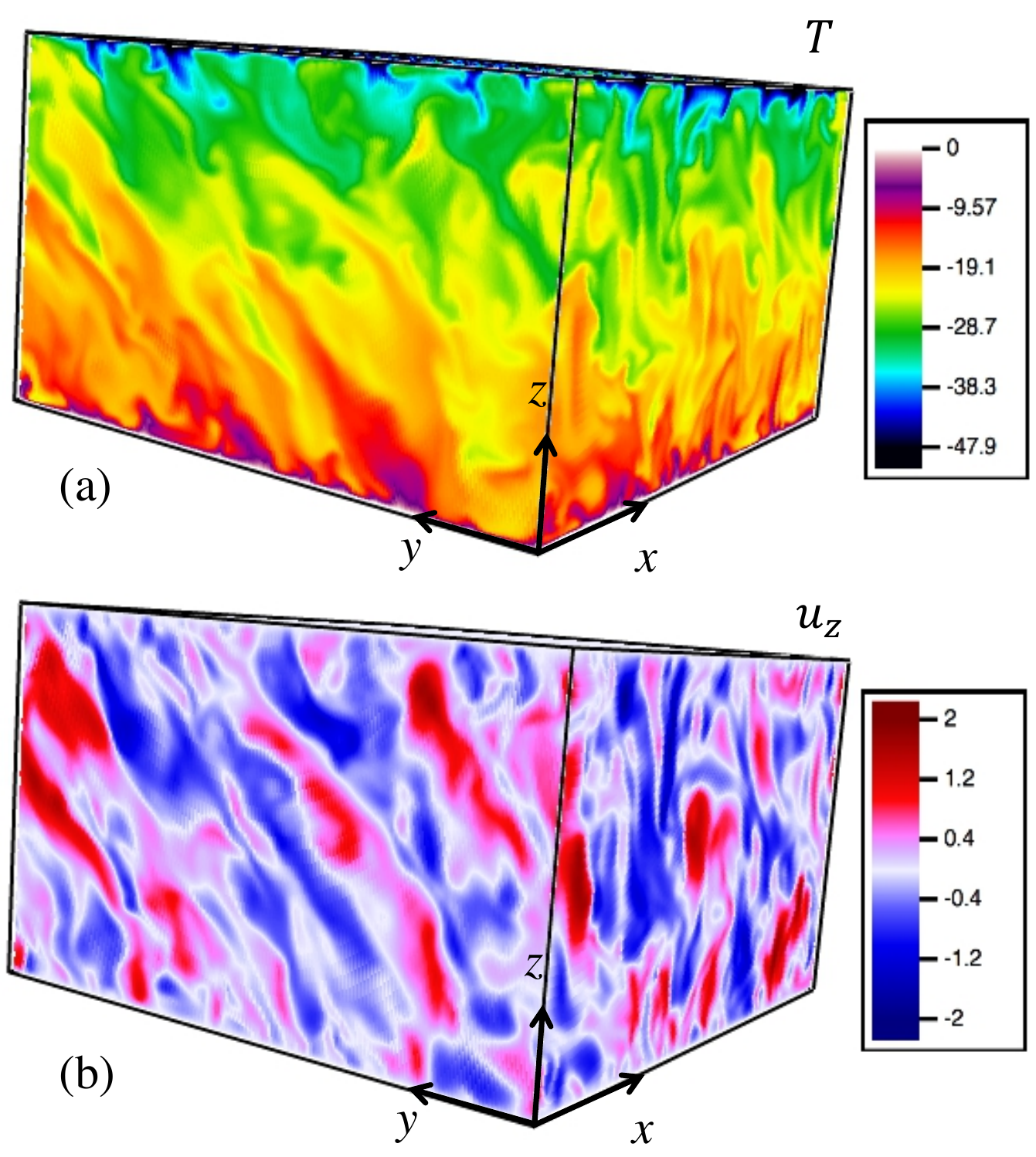}
        \caption{Snapshot of (a) $T$ and (b) $u_z$ from an illustrative simulation with $\Omega=10$ and $\phi=45^{\circ}$ (case 10A45c in Table~\ref{Table1}). The convective plumes align approximately with the rotation axis.}
    \label{plumes}
\end{figure}

We have performed a large number of rotating 3D simulations; the values of the parameters used in all of the simulations are listed in Appendix \ref{AppendixTables}, where we also report the numerical data for the bulk properties along with corresponding Rossby and vertical Reynolds numbers. The rotational influence can be seen in Fig.~\ref{plumes}, which shows the temperature and vertical velocity in an illustrative simulation, where the convective plumes are approximately aligned with the rotation axis in the $y-z$ plane. 

In this rotation-dominated regime the horizontal flow components tend to develop large-scale structures; examples of such flows are shown in Fig.~\ref{figf8} but we defer a discussion of these and their impact until section \ref{Meanflows}.

\subsection{Dependence of temperature gradient on rotation rate and latitude}\label{dTdzsims}
\begin{figure*}
\begin{center}
\includegraphics[scale=1,trim = {0mm 0mm 0mm 0mm}, clip ]{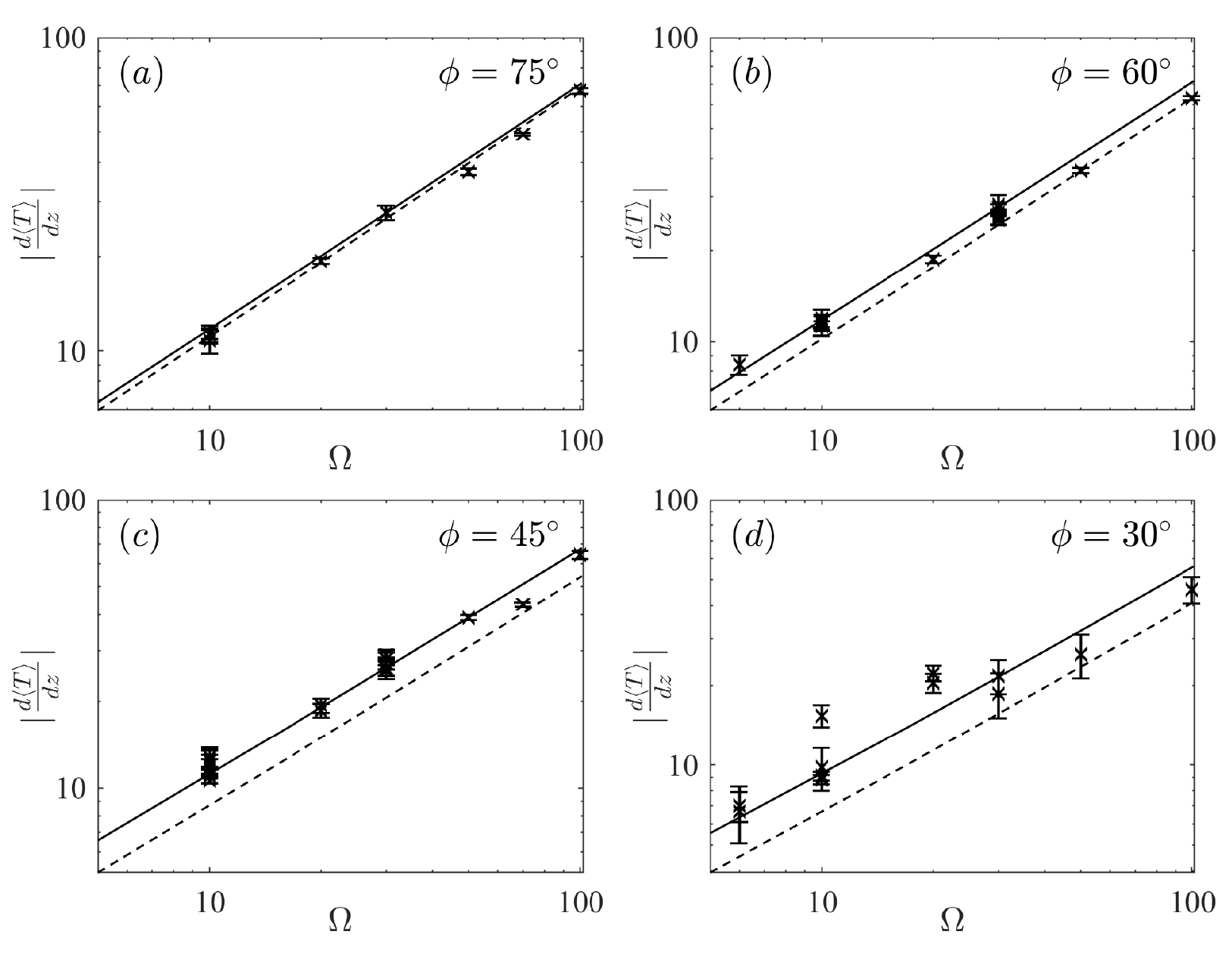}
\caption{Mean temperature gradient in the bulk (as measured over the middle one third of the convection zone) as a function of $\Omega$ for fixed latitude $\phi$ from a series of simulations from setup B (data points). The dashed lines are all proportional to $\Omega^{4/5}$ and correspond to S79's single-mode solution in the rapidly rotating limit (with $A=50$). At each latitude, the theoretical scaling of $\Omega^{4/5}$ agrees with the data quite well -- though a different proportionality constant would be needed to give better agreement with the data. The agreement with $\Omega^{4/5}$ is poorest at the smallest $\phi$. The solid lines are the theoretical predictions of multi-mode RMLT, which are computed by taking $A=0.36$ for all $\Omega$ and $\phi$ across all four panels.}\label{fig_fixphi}
\end{center}
\end{figure*}

As discussed in the introduction, we are interested in the effect of rotation on the temperature gradient at different latitudes because of its potential application to modelling stellar interiors.
Fig.~\ref{fig_fixphi} shows the variation with $\Omega$ of the mean bulk temperature gradient at four different latitudes from simulations in setup B. The mean bulk temperature gradient is calculated from the horizontally-averaged temperature profile by taking the mean gradient over the central one third of the convection zone depth.
 The associated error bars are determined by subtracting the time-averaged temperature gradient from the temperature gradient at each time and then taking the RMS average of the resulting values to give a measure of the error.

For each panel, the dashed line has a slope of $4/5$ and so it is clear that $|\frac {d \langle T \rangle}{dz}|$ scales approximately as $\Omega^{4/5}$ at fixed latitude. $|\frac {d \langle T \rangle}{dz}|\sim \Omega^{4/5}$ is the rapidly rotating limit of the predicted scaling in the RMLT of \citet{Stevenson1979}, which will be discussed further in section \ref{MLT}, and it is consistent with simulations at the poles (see e.g., Paper I). It is clear from Fig.~\ref{fig_fixphi}(d) that this scaling is not as robust for $\phi=30^{\circ}$. One possible reason for this is that, at latitudes closest to the equator, the horizontal box size in $y$ is likely too small so that the periodic boundary conditions significantly constrain the flow. As a result, some of the cases with the smallest $\phi$ are expected to disagree with the theoretical prediction of RMLT, particularly if the tilt of the modes in the $(y,z)$-plane is such that $\cot\phi\gtrsim L_y/L_z$.
For this reason we do not perform any simulations for $\phi<10^{\circ}$.

\begin{figure*}
\begin{center}
\includegraphics[scale=1,trim = {0mm 0mm 0mm 0mm}, clip ]{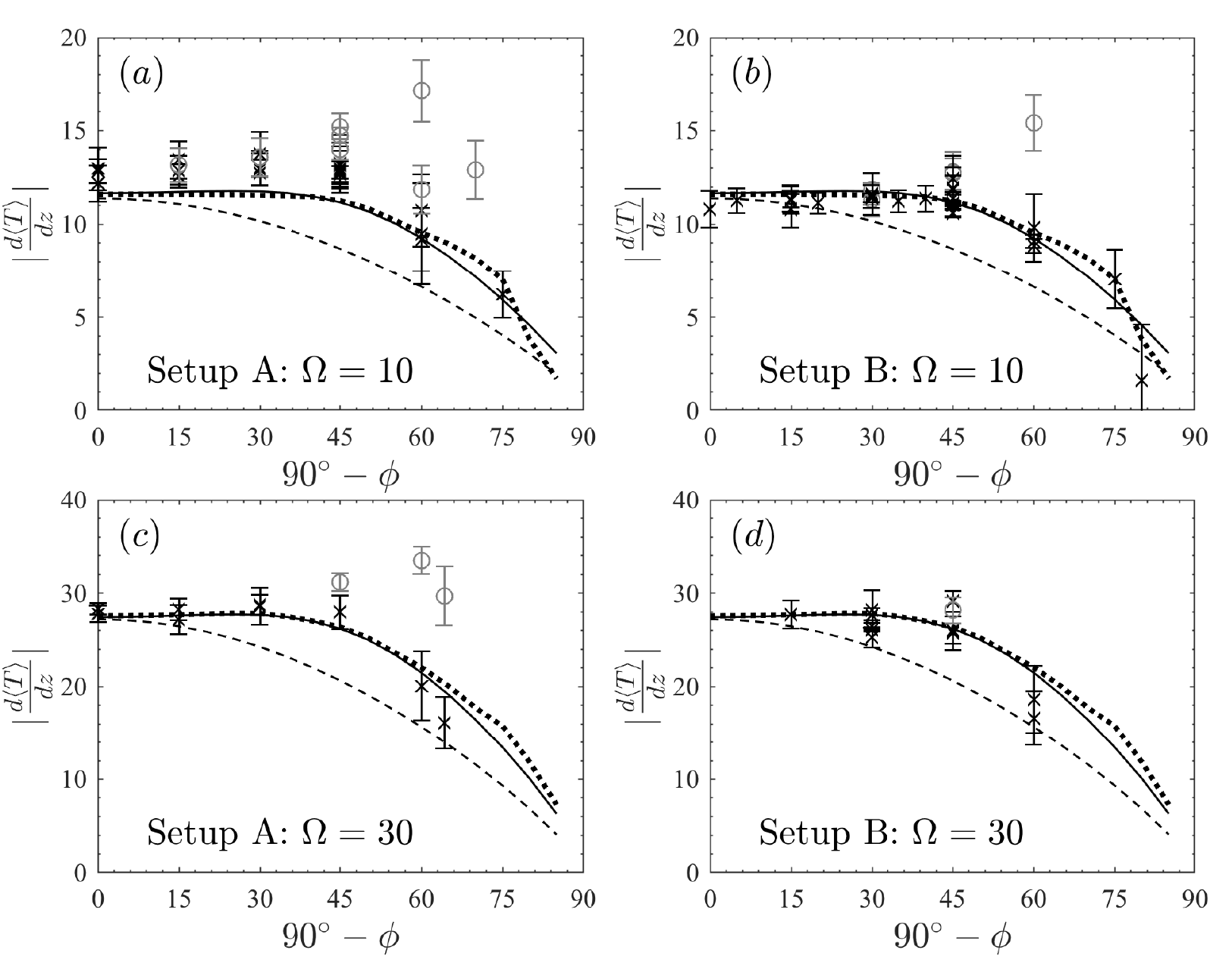}
\caption{Mean temperature gradient in the bulk (as measured over the middle one third of the convection zone) from simulations (data points) as a function of co-latitude ($90^{\circ}-\phi$) so that the pole is at the left-hand-side of the x-axis, for $\Omega=10$ (a,b) and $\Omega=30$ (c,d) in simulations using both setups (setup A, left column; setup B, right column). The dashed line gives the temperature gradient as predicted by single-mode MLT (described in appendix \ref{AppendixSingleMode}), the solid line gives the prediction from multi-mode theory (as described in section \ref{dTdzMLT}) and the dotted line gives the prediction from the restricted multi-mode theory (described in section \ref{compsim}). Again, the same proportionality constant is used for the single mode solutions ($A=50$) and for the multi-mode solutions ($A=0.36$), across all four panels. The grey circles highlight the simulations with strongest zonal flow, characterised by $\Gamma>0.1$ (see section \ref{results2}).}\label{fig_fixOm1030}
\end{center}
\end{figure*}

In a similar way, we can explore how the mean temperature gradient varies with latitude. Fig.~\ref{fig_fixOm1030} shows the dependence of the mean bulk temperature gradient on $\phi$ for two different rotation rates showing results with both setups. Note that we plot co-latitude ($90^{\circ}-\phi$) on the x-axis so that as we move along the x-axis we are moving from pole to equator.  Both sets of simulations (setups A and B) largely give the same trends, so the discussion below is relevant for both.
Whilst the trends with $\phi$ are very similar across both data sets there is a slight offset in the exact numerical values obtained. For example, $|\frac{d\langle T\rangle}{dz}|$ is consistently higher in the simulations of setup A for a fixed $\Omega$ and $\phi$. We attribute this to small differences in the depth of the convection zone owing to the different setups. The agreement in behaviour between both setups (apart from this quantitative offset) indicates that the bulk convective properties are largely insensitive to the way the convection is driven, and that the observed trends with $\phi$ (and $\Omega$) are robust.

The different data points at a fixed latitude correspond to different box sizes. We will later show in section \ref{Meanflows} that the horizontal flows that develop are strongly affected by $L_x$ and $L_y$, and that the different flows that develop (even at the same latitude) can lead to different temperature gradients.

For both $\Omega=10$ and $\Omega=30$, the general trend is that as $\phi$ is decreased from the pole, $|\frac{d\langle T\rangle}{dz}|$ remains roughly constant until $\phi\approx 45^\circ$, or perhaps slightly increases, before decreasing at smaller $\phi$. Since we are imposing a fixed flux across the domain, a larger $|\frac{d\langle T\rangle}{dz}|$ corresponds to less efficient convection. 
Naively we might expect that as we move towards the equator, the effect of rotation is reduced and so the convection is less inhibited leading to a reduced $|\frac{d\langle T\rangle}{dz}|$. However, we do not observe this effect until $\phi$ is less than $\phi=45^{\circ}$, indicating that the behaviour of $|\frac{d\langle T\rangle}{dz}|$ is more complicated than this simple expectation. Similar trends were found for $\Omega=6$ and $\Omega=20$ but for brevity we do not include them here.

\subsection{Dependence of temperature gradient on viscosity}\label{supercrit}
\begin{figure}
\begin{center}
\includegraphics[scale=1,trim = {0mm 0mm 0mm 0mm}, clip ]{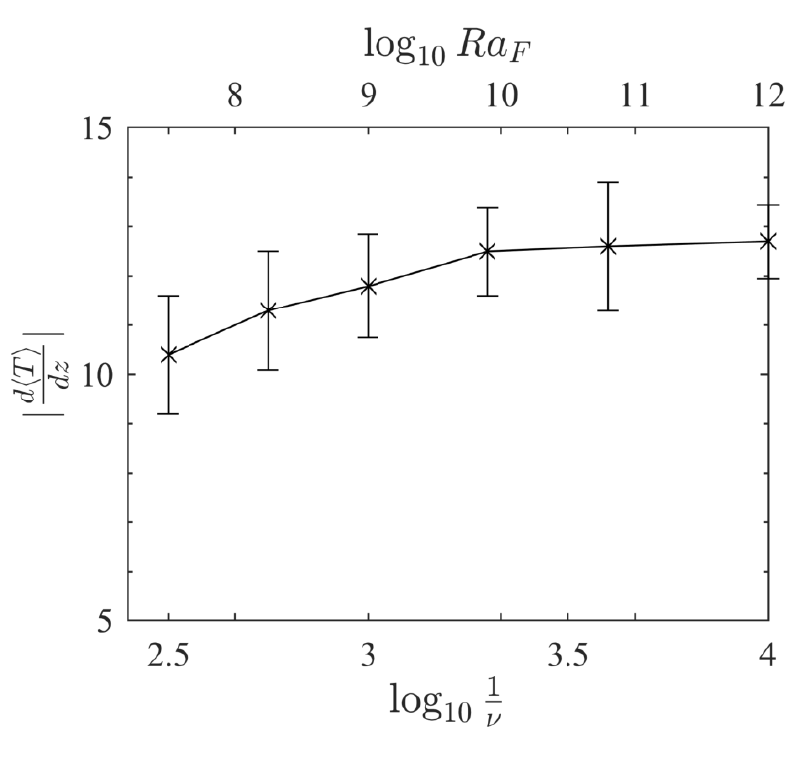}
\caption{Mean temperature gradient in the bulk of the convection zone as measured over the middle one-third of the domain as a function of $1/\nu$  (bottom axis) and $Ra_F$ (top axis) in simulations with $\Omega=10$, $\phi=45^{\circ}$ and $L_x=L_y=1.5$ done using setup B (and $F=1$). The dependence of $|d\langle T \rangle /dz|$ on $\nu$ becomes small as $\nu$ is decreased.}\label{fig_changenu}
\end{center}
\end{figure}

The viscosities and diffusivities employed here are, like those in any tractable numerical simulation, many orders of magnitude larger than the true values of these microscopic transport coefficients in stars or planets.  Equivalently, our simulations have much lower Rayleigh numbers and much higher Ekman numbers than a rapidly rotating star or planet. This will inevitably affect some aspects of the simulated flow; but if the true dynamics in stars or planets are "diffusion-free,'' as in some of the theories outlined in \S1, we may hope that these numerical effects are not too severe provided the numerical diffusion parameters are "low enough.''  Here, we therefore briefly assess the extent to which some aspects of the flows reported here -- in particular, the temperature gradient -- are influenced by these numerical parameters.  
Fig.~\ref{fig_changenu} shows how $|\frac{d\langle T\rangle}{dz}|$ changes as $\nu$ is decreased in an example set of simulations with $\Omega=10$, $\phi=45^{\circ}$, $\nu=\kappa$ and $L_x=L_y=1.5$. Clearly, the dependence on $\nu$ is reduced as $\nu$ is decreased (i.e., moving to the right in Fig.~\ref{fig_changenu}, since $1/\nu$ is plotted on the $x$-axis). Alternatively, this can be interpreted in terms of $Ra_F$ (figure \ref{fig_changenu}, top axis); the dependence of $|\frac{d\langle T\rangle}{dz}|$ on $Ra_F$ becoming reduced as $\nu$ is decreased (for fixed flux). This suggests that the simulations are approaching a regime where a diffusion-free scaling is valid. Paper I showed that bulk properties of rapidly-rotating convection in simulations at the poles are consistent with the diffusion-free predictions of RMLT.
However, we should caution that the simulation with $\nu=10^{-4}$ in Fig.~\ref{fig_changenu} has been run for long enough to obtain an equilibrated flux in the interior but large-scale flows have not fully developed. If the large-scale flows are allowed to develop then there may be changes to the value of $|\frac{d\langle T\rangle}{dz}|$ (as will be discussed in section \ref{Meanflows}). Furthermore, Fig.~\ref{fig_changenu} suggests that while some of our simulations in figures \ref{fig_fixphi} and \ref{fig_fixOm1030} are not quite in a regime where viscosity is entirely negligible, viscous effects appear to be weak.

It must be noted that the range of parameters probed by our simulations is necessarily still limited by numerical resources.
In rotating convection, the critical Rayleigh number (the value of Rayleigh number at which convection onsets) scales with $Ek^{-4/3}$ \citep{Chandra1961} and hence depends on $\nu$, $\Omega$ and $\phi$ (and well as $H$). Therefore, while decreasing $\nu=\kappa$ by a factor of $a$, say, (for fixed $F, H$) increases $Ra_F$ by a factor of $a^3$, the supercriticality of the convection (if defined as the ratio of $Ra_F$ to the onset value) is, by comparison, only moderately increased by a factor of $a^{5/3}$ (since  decreasing $\nu$ also decreases $Ek$ and hence increases the critical Rayleigh number). 
Thus, to obtain a large range of supercriticalities in simulations with fixed $F$ and $\Omega$ is relatively challenging. 
Indeed, the majority of our calculations possess a Rayleigh number of up to approximately 10 times the critical Rayleigh number at each rotation rate; the cases sampled in Fig.~\ref{fig_changenu} probe a larger range (including approximately 20 times critical in the most extreme case) but it was not practical to repeat this at all rotation rates and latitudes. We have also performed a subset of simulations varying $Ra_F$ at fixed $Ek$ to obtain a larger range of supercriticalities (see section \ref{sec:connection} and Fig.~\ref{figNuRaF}).

\section{Rotating Mixing Length Theory}\label{MLT}
\subsection{Temperature gradient from MLT}\label{dTdzMLT}
We seek to understand physically the behaviour of $|\frac{d\langle T\rangle}{dz}|$ described in section \ref{sims} using the framework of RMLT.
Paper I showed that the RMLT of S79 agrees very well with simulations at the pole and here we investigate whether it can explain the variation of $|\frac{d\langle T\rangle}{dz}|$ at other latitudes.

Following S79 (see also \citealt{FG78} and Paper I), we present the arguments for rotating MLT as a possible physical explanation for the results obtained with our numerical simulations. We begin with the (Boussinesq) linearised equations for perturbations to a linear background temperature profile $T_B=T_0-N_{*}^2 z$ (where the buoyancy frequency $N^2=-N^2_*$), in the absence of viscosity and thermal diffusion:
\begin{eqnarray}
&&\frac{\partial \bm u}{\partial t} + 2\bm\Omega \times \bm u = - \nabla p + T\bm e_z, \label{linmom}\\
&&\frac{\partial T}{\partial t} = N_{*}^2 u_z,\label{lintemp}\\
&&\nabla \cdot \bm u =0.\label{linincomp}
\end{eqnarray}
The symbols are the same as in (\ref{momeq})-(\ref{incompeq}) except that $T$ is now the perturbation to a linear background profile (as opposed to the total temperature).
We adopt impenetrable boundary conditions, i.e., $u_z=0$ on $z=0$ and 1. Since equations (\ref{linmom})-(\ref{linincomp}) are linear, we may seek growing modes of the form (\citealt{H79})
\begin{eqnarray*}
u_z\vert_{\boldsymbol k} =\mathrm{Re}\left[\hat{u}_z(\boldsymbol{k}) \exp\left(\mathrm{i}\left(k_x x + k_y y -k_y\tau z\right)+\sigma t\right)\sin n \pi z\right],
\end{eqnarray*}
where $\tau=4\Omega_z\Omega_y/(4\Omega_z^2+\sigma^2)$, and similarly for other variables such as $T$, and $\boldsymbol{k}=(k_x,k_y,n\pi)$. This allows us to obtain the following dispersion relation for the growth rate\footnote{This gives the same growth rate as S79 Eq.~33 (but his Eq.39 is incorrect since it omits a term) and \citet{FG78} Eq.~2.33 (once we correct a typo in their term involving $l^2$).
}, $\sigma$ 
\begin{eqnarray}
\nonumber
&& \sigma^4+\left(4\Omega_z^2(2n^2\pi^2+k_{\perp}^2)+4\Omega_y^2 k_y^2-N^2_{*}k_\perp^2\right) \frac{\sigma^2}{k^2} \\
&& \hspace{1in} + \frac{4\Omega_z^2}{k^2}\left(4\Omega_z^2n^2\pi^2-N^2_{*}k_\perp^2\right)=0,
\label{growth}
\end{eqnarray}
where $k^2=k_\perp^2+n^2\pi^2$, and $k_{\perp}^2=k_x^2+k_y^2$.

 There are two arguments required to formulate RMLT for a single mode. We first relate the convective heat flux to the velocity, and then we relate the velocity to the linear growth rate by assuming that the latter balances a nonlinear cascade rate. Finally, we sum up over all of the modes to obtain a multi-mode RMLT. 

The convective heat flux due to a single mode is given by 
\begin{eqnarray}
\label{fluxdef1}
F\vert_{\boldsymbol{k}} = \langle u_z\vert_{\boldsymbol{k}} \  T\vert_{\boldsymbol{k}} \rangle
  \propto \hat{u}_{z}(\boldsymbol{k})\hat{T}^* (\boldsymbol{k}) + \mathrm{c.c.}, 
\nonumber
\end{eqnarray}
where $\langle \cdot\rangle$ denotes a spatial average and $\mathrm{c.c.}$ represents the complex conjugate. By multiplying equation (\ref{linmom}) by the velocity amplitude of a mode, $\boldsymbol{u}\vert_{\boldsymbol{k}}$, and spatially averaging, we can show that  (e.g.~S79)
\begin{eqnarray}
\label{fluxdef2}
   F\vert_{\boldsymbol{k}} = \sigma \langle |\boldsymbol{u}\vert_{\boldsymbol{k}}|^2\rangle. 
    \nonumber
\end{eqnarray}
This implies that
the typical flux $F_\lambda$ on a characteristic lengthscale $\lambda$ satisfies 
\begin{eqnarray}
\label{fluxdef3}
F_\lambda \sim \sigma u_\lambda^2,
\end{eqnarray}
where $u_\lambda$ is a typical value of the velocity magnitude on a lengthscale $\lambda$. (If one wished to be
more precise, one could define $u_{\boldsymbol{\lambda}}$ as the structure function $u_{\boldsymbol{\lambda}}^2 = \langle |\boldsymbol{u}(\boldsymbol{x}+\boldsymbol{\lambda})-\boldsymbol{u}(\boldsymbol{x})|^2\rangle\sim 
k^3 |\hat{\boldsymbol{u}}(\boldsymbol{k})|^2/V$, where $V$ is the volume of the domain.)

We next suppose that the amplitude of each convective mode is controlled by the requirement that its growth rate balances its nonlinear cascade rate, i.e.,
\begin{eqnarray}
\label{amp}
\sigma(\boldsymbol{k})\sim k_{tot} \, u_\lambda.
\end{eqnarray}
Here we have defined $k_{tot}^2=k_{\perp}^2+(n\pi\pm k_y\tau)^2\approx k_{\perp}^2+k_y^2\tau^2$ since $k_y\tau\gg k_z=n\pi$ for many of the modes in our simulations. This accounts for the tilted nature of the convective modes when $\phi\ne90^{\circ}$. The contribution to the heat flux on a lengthscale $\lambda$ is then obtained by substituting Eq.~\ref{amp} into Eq.~\ref{fluxdef3} to obtain
\begin{eqnarray}
\label{Fmode}
    F_{\lambda} \sim\frac{\sigma^3}{k_{tot}^2}.
\end{eqnarray}

The total heat flux is the weighted sum of this quantity over all of the modes, i.e.,
\begin{eqnarray}
F = \iiint F_\lambda \mathrm{d}\ln k_x \mathrm{d}\ln k_y \mathrm{d} \ln n,
\label{Flux}
\end{eqnarray}
where we set 
\begin{eqnarray}\label{Amode}
    F_{\lambda}=A\frac{\sigma^3}{k_{tot}^2},
\end{eqnarray}
and $A$ is a constant to be determined from simulations. 
This integral takes into account the relative volume of $k$-space occupied by modes with a given $F_\lambda$. The reason for the $\ln$ factors is because $F_\lambda$ represents the typical value of the flux on a lengthscale $\lambda$, which differs from the flux due to Fourier modes with a given $\boldsymbol{k}$.

Our picture is that the convection is dominated by a sea of uncorrelated modes whose amplitudes are each determined by Eq.~\ref{amp}. The amplitude-limiting criterion (Eq.~\ref{amp}) is a highly simplified model of nonlinear effects, but we will later show that it is appropriate for explaining the bulk properties of rotating convection. This picture is not appropriate for the modes with short enough lengthscales such that viscosity and thermal diffusion are important.

If we know the total flux $F$, Eq.~\ref{Flux} can be viewed as an inverse problem to determine the required value of $N_*^2$ to transport this heat for a given $H$, $\Omega$ and $\phi$, which we can solve numerically (e.g., this is straightforward to accomplish using fsolve in Matlab). We will refer to this formalism as the multi-mode approach since the integral in (\ref{Flux}) is over many modes. A simpler way to proceed is to suppose that the heat transport is dominated by a single mode, namely the one that transports the most heat, as considered by S79 and Paper I (the details of this approach and its relation to the arguments in Paper I are given in Appendix \ref{AppendixSingleMode}); we will refer to this simpler formalism as the single-mode approach. In cases with rotation and gravity aligned (i.e., at the poles), Paper I found the single-mode theory matched the numerical results very well. However, (as will be shown below), this theory does not describe as well the dynamics in non-polar regions. This motivates us to retain multiple modes (strictly all of the modes under the above assumptions) in our analysis below.

\begin{figure*}
\includegraphics[scale=1,trim = {0mm 0mm 0mm 0mm}, clip ]{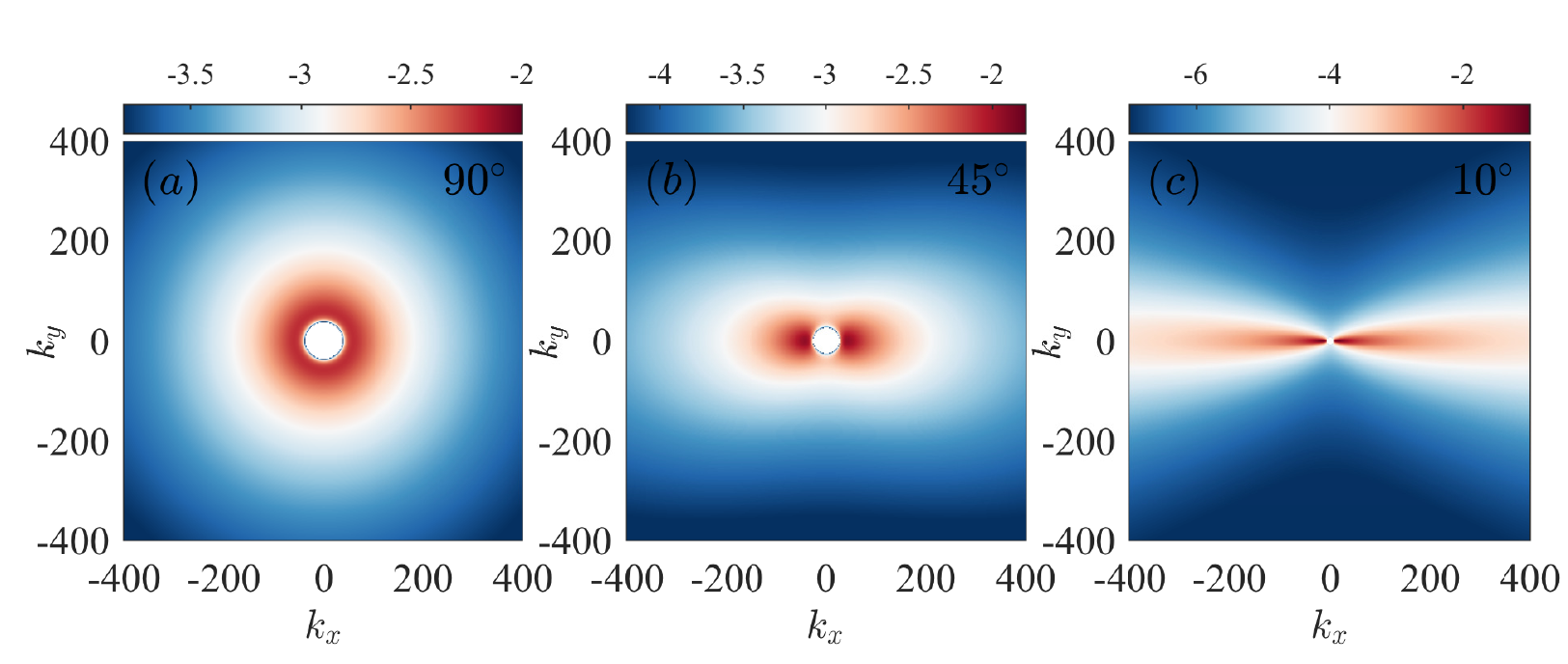}
\caption{Logarithm of the single mode flux, i.e.~$\log_{10}F_\lambda=\log_{10}(A\frac{\sigma^3}{k^2})$, as a function of $k_x$ and $k_y$ for $\Omega=30$ and three different latitudes ($\phi=90^{\circ}, 45^{\circ}$ and $10^{\circ}$). We fix $A=0.36$. This shows that modes with $k_y\sim 0$ dominate the transport at non-polar latitudes, and that the modes that maximise $F_\lambda$ occupy a shrinking volume of $k$-space as $\phi\rightarrow 0^{\circ}$.} \label{figf1:mickeymouse}
\end{figure*}

To see why including multiple modes might be necessary as we move away from the polar regions, Fig.~\ref{figf1:mickeymouse} displays $F_\lambda$ on the $(k_x,k_y)$-plane for modes with $n=1$, with $\Omega=30$ for $\phi=90^{\circ}$ (pole), $45^{\circ}$ and $10^{\circ}$. This shows that the modes that transport heat most efficiently are those with $k_y=0$ whenever we are not at the poles. In addition, the contours of $F_\lambda$ become less symmetric as $\phi\rightarrow 0^{\circ}$, implying that the single mode that maximises $F_\lambda$ is no longer representative, unlike in the polar case where the contours of $F_\lambda$ are symmetric. 
We will show that this modifies the predictions of RMLT.
Note that we only include modes where $F_\lambda$ is positive; since $\sigma$ is negative for $k_x$ and $k_y$ both small \citep{H79} there exists a void region at the centre of each plot in Fig.~\ref{figf1:mickeymouse}.

To form the multi-mode predictions, we assume that $n=1$ dominates, based on the single-mode result in Appendix B, but allow any $k_x$ and $k_y$ value. We fix the arbitrary normalisation constant, $A$, in the theory by matching the value of $N_*^2$ at the pole to the value obtained in simulations \textit{for one chosen value} of $\Omega$. The theory then \textit{predicts} the $\Omega$ and $\phi$ dependence of $N_*^2$. In practice, we obtain this prediction by summing up the modes numerically on a discrete grid of $k_x$ and $k_y$ values such that the integral in Eq.~\ref{Flux} is converged, and then find the value of $N_*^2$ that gives our desired total flux $F$ for each $\Omega$ and $\phi$. Note, we only include convectively unstable modes in our calculations, i.e., we set $F_{\lambda}=0$ if $Re(\sigma)\leq 0$.

\subsection{Other predictions from MLT}\label{otherMLT}
Once we have obtained $N_*^2$, we may also obtain expressions for the corresponding RMS velocity ($u_{z,rms}$) and temperature fluctuations ($\delta T_{rms}$) as follows.
We assume that there is a good correlation between rising warm fluid and falling cool fluid, so that the heat flux carried by modes with these wavenumbers is given by
\begin{eqnarray}\label{Bdef}
F_{\lambda} \sim  v_z \delta T,
\end{eqnarray}
where $v_z$ and $\delta T$ are the RMS vertical velocity and temperature fluctuations for these modes.
Equation (\ref{lintemp}) gives
\begin{equation}
\label{MLTheat}
    \sigma \delta T = N_*^2v_z,
\end{equation}
then combining (\ref{Amode}), (\ref{Bdef})  and (\ref{MLTheat}) gives 
$v_z\sim \sqrt{A}\sigma^2/k_{tot}N_*$ (e.g.~S79), and so
\begin{eqnarray}
u_{z,rms} \sim \sqrt{\iint \frac{\sigma(\boldsymbol{k})^4}{k_{tot}^2N_*^2} \,\mathrm{d}\ln k_x \mathrm{d}\ln k_y}. 
\label{uzint}
\end{eqnarray}
(\ref{MLTheat}) then gives $\delta T = N_*^2 v_z/\sigma \sim \sqrt{A}N_*\sigma/k_{tot}$,
and so
\begin{eqnarray}
\delta T_{rms} \sim \sqrt{\iint \frac{N_*^2\sigma^2}{k_{tot}^2}\,\mathrm{d}\ln k_x \mathrm{d}\ln k_y.}
\label{delTint}
\end{eqnarray}

Scalings for the dominant wavenumbers of the convection can be obtained in the following way:
\begin{eqnarray}\label{khat}
\hat{k} = \frac{\int k_\perp F_\perp\,\mathrm{d} k_\perp}{\int F_\perp \mathrm{d} k_\perp}
\end{eqnarray}
where $k_\perp=(k_x^2+k_y^2)^{\frac{1}{2}}$ and $F_\perp=F_{\lambda}/{k_xk_y}$. To obtain a smoother profile, $F_\perp$ is binned in integer bins of $k_\perp$.

The $\Omega$ and $\phi$ variation of each quantity is a meaningful prediction of the theory as is the anisotropy in the $x$ and $y$ directions.


\subsection{Comparison with simulations}\label{compsim}
\subsubsection{Comparison of temperature gradient}
Returning first to Fig.~\ref{fig_fixphi}, which shows the variation with $\Omega$ of the mean bulk temperature gradient at four latitudes, we focus now on the over-plotted solid lines.  These show the prediction of the multi-mode theory for $|d\langle T\rangle /dz|$ at each latitude and rotation rate.  It is clear that the scaling with rotation rate in this theory is virtually identical to the $\Omega^{4/5}$ scaling implied by the single-mode theory in the rapidly-rotating limit, and hence that our simulations are qualitatively in agreement with this particular prediction of both multi-mode and single-mode RMLT. However, the multi-mode theory provides better quantitative agreement with the data when the same proportionality constant, $A=0.36$, is used at all latitudes; the single-mode prediction (with $A=50$) becomes progressively less appropriate at latitudes closer to the equator.

Fig.~\ref{fig_fixOm1030} shows how both the theoretical and the simulation values of $N_*^2=|\frac{ {\rm d}\langle T \rangle}{{\rm d}z}|$ vary as a function of co-latitude ($90^{\circ}-\phi$) for the simulations of setup A (left-hand column) and in the simulations of setup B (right-hand column), for $\Omega=10$ and $\Omega=30$ and a range of horizontal box sizes.
The dashed lines show how the temperature gradient would be expected to scale with $\phi$ based on S79's (single-mode) theory, which only depends on $\Omega_z$ (the vertical component of $\boldsymbol{\Omega}$),
whilst the solid line shows how the temperature gradient varies as determined by the multi-mode theory described in section \ref{dTdzMLT}.

For both sets of simulations the multi-mode prediction gives better agreement than the single-mode prediction with the simulation data in that $|\frac{d\langle T\rangle}{dz}|$ stays flatter until smaller $\phi$ as we move away from the poles. Indeed, the multi-mode prediction does a reasonably good job of predicting the dependence of the mid-layer temperature gradient on $\phi$ for all $\Omega$ considered here.
There are however a few outliers, which lie noticeably above the solid line: these are 
simulations with horizontal box sizes such that strong zonal flows have developed, which we will discuss further in \S~\ref{Meanflows} (note the cases with strongest zonal flows are highlighted with a grey circle).
The poorest agreement between the theoretical lines and the simulations occurs at the latitudes closest to the equator, which as discussed previously could also be a result of the constraining effects of the horizontal periodic boundaries in cases with $L_y$ being too small.

The multi-mode prediction is obtained by summing up over enough modes that convergence is obtained (i.e., the integral is converged and each result does not vary when higher wavenumber modes are added). Since the modes in a numerical simulation are determined by the box size and the resolution, we also perform calculations which include only the discrete set of $k_x$ and $k_y$ values that are present in the simulation (using a typical resolution and box size for each set of parameters); this leads to non-smooth predictions (dotted lines). These dotted lines roughly follow the full multi-mode prediction indicating that in almost all cases the discreteness of the modes in our simulation is unlikely to cause a significant departure of our numerical data from the multi-mode theoretical prediction, but some differences arise because of the different modes included in each calculation.

\subsubsection{Comparison of the heat flux spectra}\label{compspectra}
\begin{figure*}
\begin{center}
\includegraphics[scale=1,trim = {0mm 0mm 0mm 0mm}, clip ]{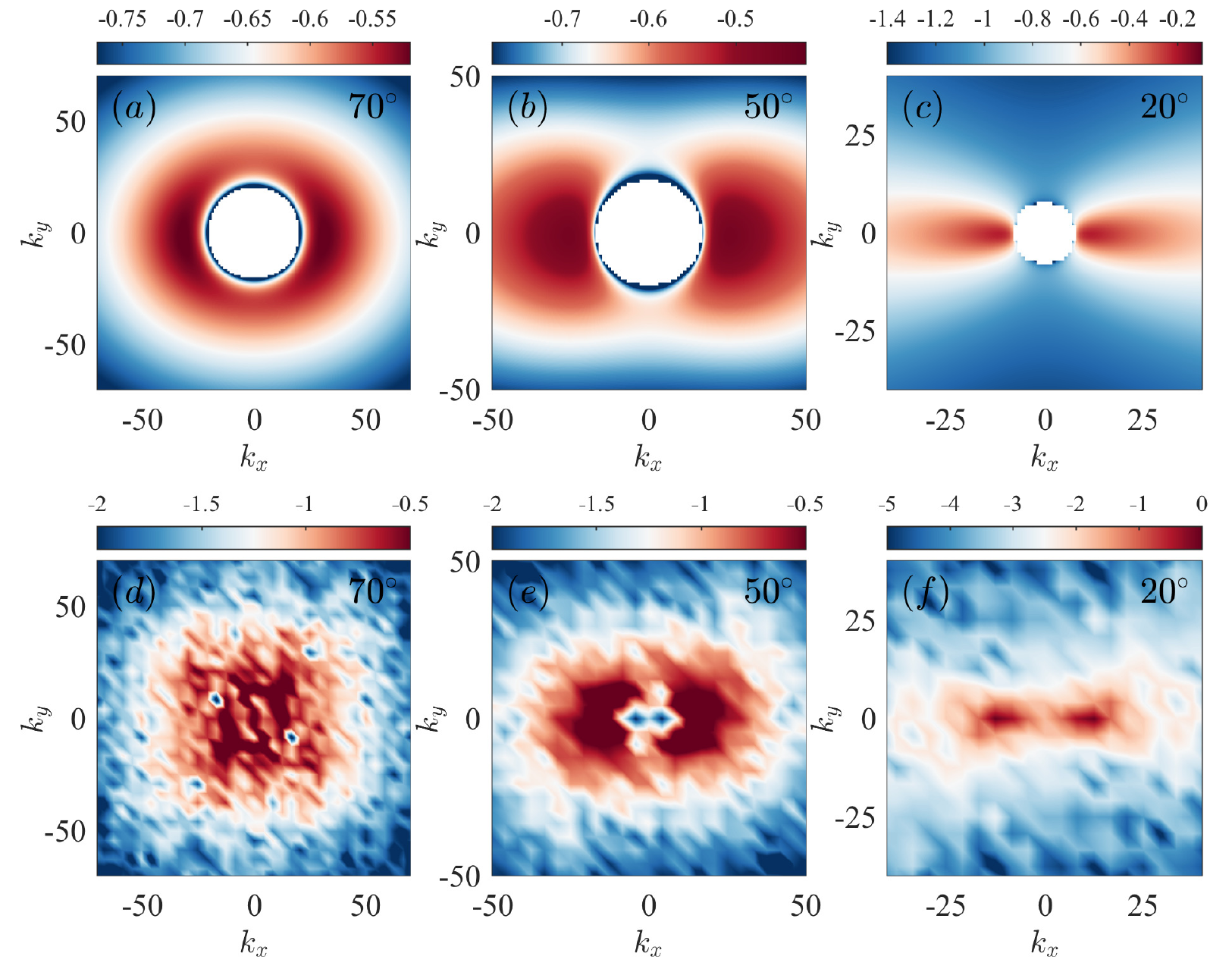}
\caption{Comparison of the theoretical and simulation heat flux spectra. Logarithm of the single mode flux, i.e.~$\log_{10}F_\lambda\sim\log_{10}(A\frac{\sigma^3}{k^2})$, as a function of $k_x$ and $k_y$ for $\Omega=20$ and three different latitudes ($\phi=70^{\circ}, 50^{\circ}$ and $20^{\circ}$) is given in (a-c). The equivalent heat flux spectrum from a numerical simulation is given for each latitude: case 20B70a in (d), case 20B50a in (e) and case 20B20a in (f). Note that the theoretical expression has been plotted with $A=0.36$ and and the colorbars have been adjusted by eye to emphasise key features.}\label{figf5a}
\end{center}
\end{figure*}

We can also compare qualitatively the theoretical $F_\lambda$ as a function of $k_x$ and $k_y$ with the heat flux spectrum obtained from simulations. We obtain the values from simulations by using data at the mid-plane $z=L_z/2$ for $u_z(x,y,L_z/2,t)$ and $T(x,y,L_z/2,t)$, and we compute the spatial discrete Fourier transform of this data to produce $\hat{u}_{z,k}(k_x,k_y,t)$ and $\hat{T}_k(k_x,k_y,t)$. These quantities are used to compute the heat transport spectrum $\hat{u}_{z,k}(k_x,k_y,t)\hat{T}^*_k(k_x,k_y,t)$. 
In Fig.~\ref{figf5a} we compare $F_\lambda$ (top row) with $Re(\hat{u}_{z,k} \hat{T}^*_k)$ (bottom row) for $\Omega=20$, at three different latitudes. Note that these quantities are not expected to match quantitatively due to their different definitions. In particular, $F_\lambda$ represents the heat flux in a logarithmic interval in $k$-space, whereas $\hat{u}_{z,k}(k_x,k_y,t)\hat{T}^*_k(k_x,k_y,t)$ represents the heat flux in a unit interval in $k_x$ and $k_y$. However, both quantities represent a measure of the heat transport for the modes, so they should share common features, and we expect their azimuthal structures to be similar in the $(k_x,k_y)$-plane.

We show the time-averaged spectra from a number of snapshots of the data at different times in the turbulent state to reduce noise, though turbulent fluctuations have not been eliminated entirely. In each case the logarithm of the flux is plotted and the quantities have been scaled by a constant to aid comparison. 
We have only plotted the low wavenumber modes because RMLT is not expected to be valid for high wavenumbers, where rotation should be unimportant and diffusion should be the dominant process. But at small $k$, if RMLT is correct it should approximately describe the data. However, we should note that RMLT does not capture all possible nonlinear interactions, and in particular omits transfers of energy into `stable' modes at small $k$ for which $\sigma(\boldsymbol{k})<0$. Nonlinear interactions between `unstable modes' and `stable modes' are of course possible in the simulations, so we may expect a departure for the smallest wavenumbers from RMLT. Nevertheless, if RMLT is correct it should approximately capture the shape of the contours, including the anisotropy between $k_x$ and $k_y$.

Overall, there is very good qualitative agreement in Fig.~\ref{figf5a} between the simulations and the theory; for the simulation closest to the pole, the spectra should be close to being symmetric about the origin, with little differences along the $k_x$ and $k_y$ axes, and this is essentially what we observe. Furthermore, the asymmetry between the $x$ and $y$ directions (as shown previously in Fig.~\ref{figf1:mickeymouse}) increases as we move towards the equator, in both simulations and theory. 
The biggest discrepancies between the theory and simulations occurs in these plots at the smallest $k_x$ and $k_y$, which is probably because the theory 
neglects nonlinear interactions that are present in reality.

\begin{figure*}
\begin{center}
\includegraphics[scale=1,trim = {0mm 0mm 0mm 0mm}, clip ]{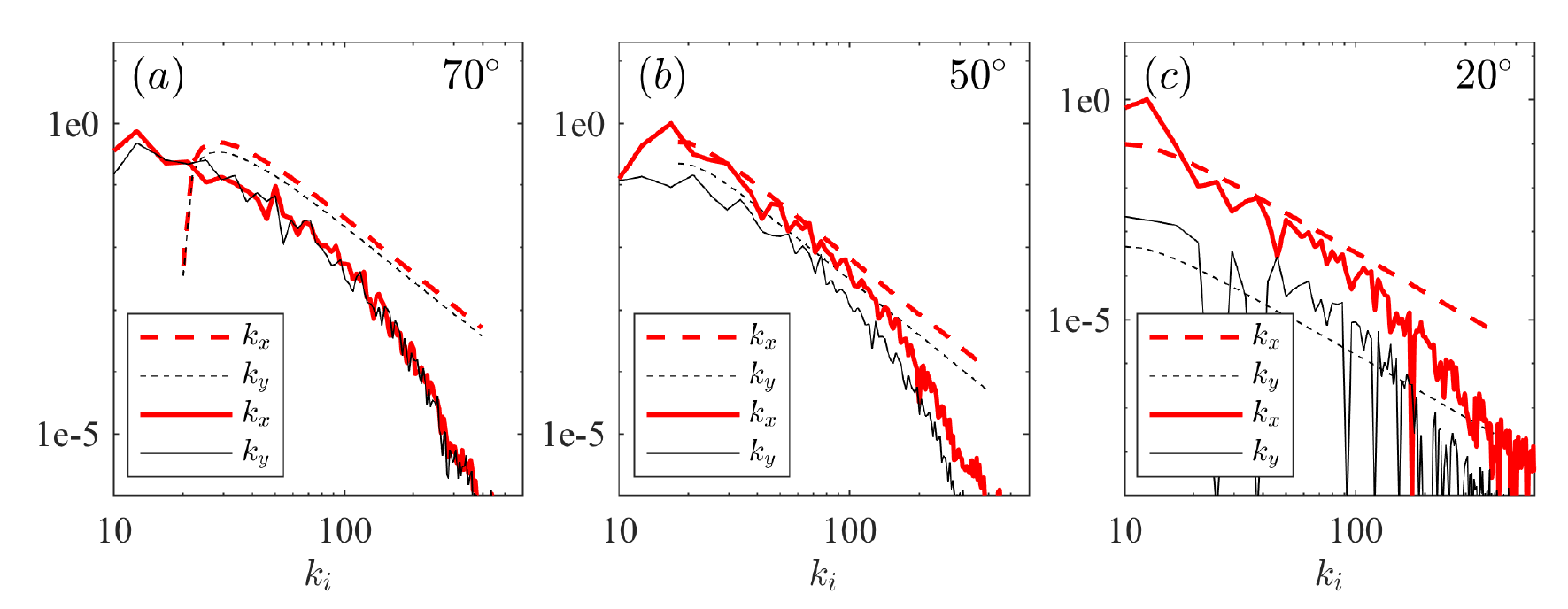}
\caption{Comparison of the theoretical and simulation 1D heat flux spectra. The solid lines give the heat flux spectrum along the $k_x$ axis (thick, red lines) and along the $k_y$ axis (thin, black lines) and the dashed lines give the RMLT predictions $F_\lambda/k_x$ (thick, red lines) and $F_\lambda/k_y$ (thin, black lines). In all cases $\Omega=20$ and in (a) $\phi=70^{\circ}$, in (b) $\phi=50^{\circ}$ and in (c) $\phi=20^{\circ}$. The simulations used are give in Table~\ref{Table1} in cases 20B70a (a), 20B50a (b) and 20B20a (c).  The slope of the spectra at intermediate wavenumbers is well described by the theory and the anisotropy between the $k_x$ and $k_y$ is also well captured by the theory, though the agreement is poorest at the smallest $\phi$ plotted where box size effects are likely to be important.}\label{figf6}
\end{center}
\end{figure*}

An alternative comparison between theory and simulations is by plotting the heat flux spectrum along the $k_x$- and $k_y$-axes from simulations and comparing this with the RMLT predictions $F_\lambda/k_x$ and $F_\lambda/k_y$, respectively (these factors are approximate and arise from integration over $\mathrm{d}\ln k_x \mathrm{d}\ln k_y$ in $F_\lambda$, versus $\mathrm{d} k_x \mathrm{d} k_y$ for $Re(\hat{u}_{z,k} \hat{T}^*_k)$). The results for $\Omega=20$ are shown in Fig.~\ref{figf6}. We have arbitrarily scaled both theoretical lines together, by the same normalisation factor, as well as that of the simulation data, so that both sets of data have the same magnitude at a particular $k_x$ or $k_y$. The overall magnitudes are therefore arbitrary, but the shape and any differences between the lines along the $k_x$- and $k_y$-axes are meaningful tests of the theory.

We again see the symmetry breaking that occurs as we move away from the pole. Note that linear theory predicts $\sigma\sim N_*$ for the fastest growing mode, approximately independent of $k$. As a result, RMLT predicts $F_\lambda\sim 1/k^2$, so that $F_\lambda/k_x\sim 1/k_x^3$ and $F_\lambda/k_y\sim 1/k_y^3$. The slopes of the 1D spectra are well approximated by these theoretical predictions at intermediate scales. The asymmetry between $k_x$ and $k_y$ is accurately captured at all three latitudes considered. This is particularly clear from the data with $\phi=20^{\circ}$, where the heat transported by modes with $k_y=0$ is more than an order of magnitude greater than that by modes with $k_x=0$.

Overall, we have shown that the spectra in our simulations are quite well described by RMLT for the wavenumbers where we expect this to apply.

\subsubsection{Comparison of other quantities}\label{subsec:compOQ}
Other predictions from RMLT obtained in \S\ref{otherMLT} can also be tested against simulations. 
In order to test these we calculate from the simulations, the RMS values of $u_z$ and $T'=T-\langle T \rangle$ at the mid-plane of the box. These values are then time-averaged over the duration of the simulation in the turbulent quasi-steady state.
The associated errors are estimated by subtracting the time-averaged RMS quantity ($u_z$ or $T'$) from the quantity at each time before taking a RMS average of the resulting values to give an overall "error".
To obtain $\hat k$ from the simulations we use data at the midplane ($z=L_z/2$) for $u_z$ and $T$ for a selection of data at various times in the simulation (Nek5000 data is then interpolated onto a uniform $xy$-grid), which is then used to calculate $\hat{u}_{z,k}(k_x,k_y,t)\hat{T}^*_k(k_x,k_y,t)$ as described in section \ref{compspectra}. We can the calculate the equivalent integral to that in (\ref{khat}) by replacing $F_{\perp}$ with $Re(\hat u_{z,k}\hat T_k^*)$, i.e.,
\begin{equation}\label{khatsims}
  \hat{k} = \frac{ \int k_{\perp} Re(\hat{u}_z\hat{T}^*) d k_{\perp} }{ \int Re(\hat{u}_z\hat{T}^*) d k_{\perp}},
\end{equation}
where $k_{\perp}=(k_x^2+k_y^2)^{1/2}$ bins have been used. 

\begin{figure*}
\begin{center}
\includegraphics[scale=1,trim = {0mm 0mm 0mm 0mm}, clip ]{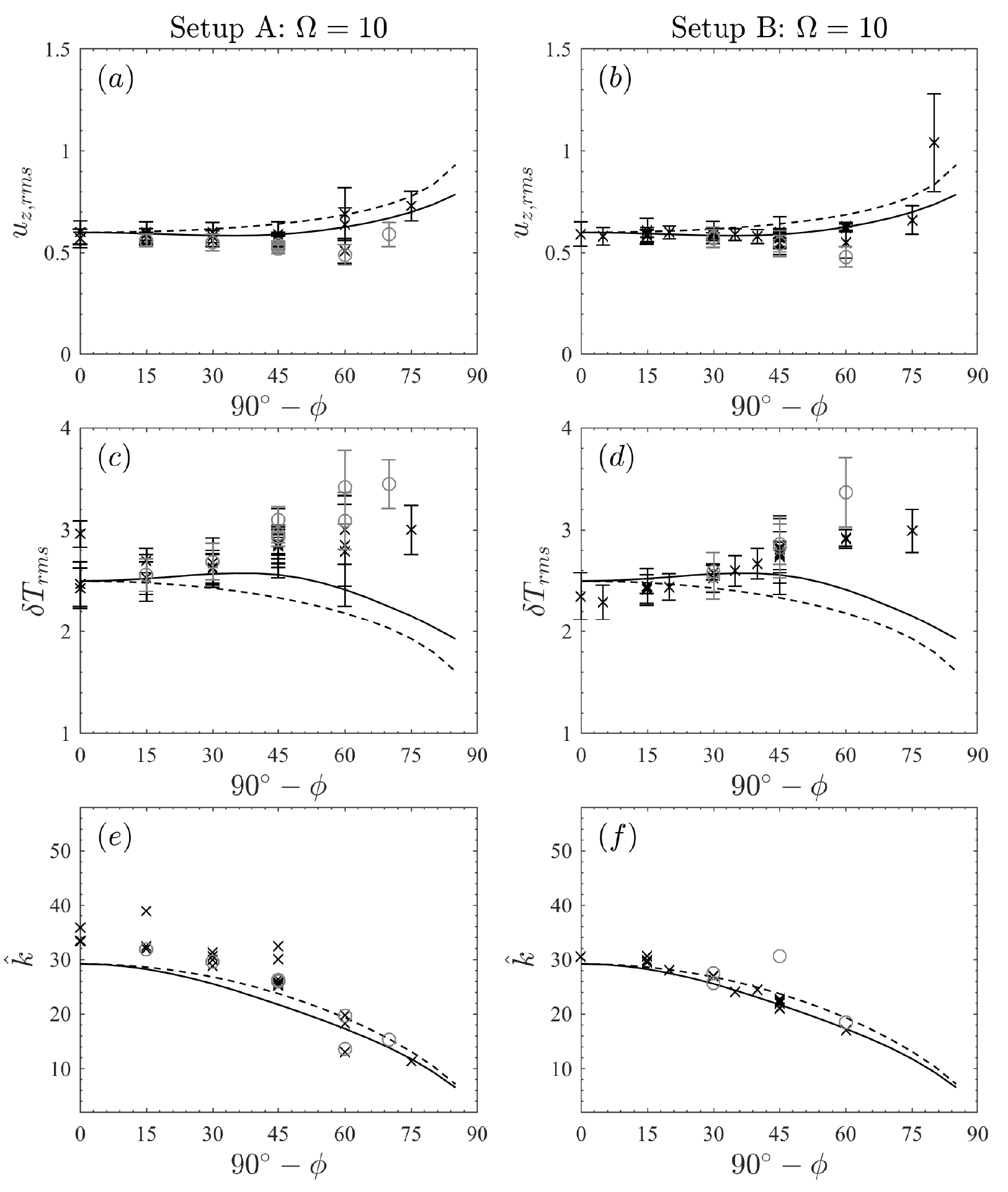}
\caption{RMS vertical velocity (a), (b); RMS temperature fluctuation (c), (d) and "dominant" wavenumber $\hat{k}$ at the midplane for simulations with $\Omega=10$ and varying $\phi$ using setup A (1st column) and setup B (second column). The symbols represent data from simulations, the solid lines are predictions from multi-mode theory (with $A=0.36$) and the dashed lines are from single-mode theory (with $A=50$). The grey circles highlight the simulations with strongest zonal flow, characterised by $\Gamma>0.1$ (see section \ref{results2}). }\label{fig_fixOm10_other}
\end{center}
\end{figure*}

The comparison between theory and simulations for $\Omega=10$ is shown for both setups in Fig.~\ref{fig_fixOm10_other}.
Again, the trends for both setups are similar reinforcing the robustness of our results and permitting the same description for both data sets.
Note the expressions in (\ref{uzint}) and (\ref{delTint}) give the predicted trend with $\phi$ but not the precise magnitude and so we normalise both the single- and multi-mode predictions such that they agree with the simulation at the pole. 
$u_{z,rms}$ does not vary significantly until $90^{\circ}-\phi\gtrsim 75^{\circ}$ and this behaviour is captured by both the single- and the multi-mode theoretical predictions.
For the smallest $\phi$, the increase of $u_{z,rms}$ with $90^{\circ}-\phi$ in the data is not well described by the theoretical predictions; again, this could be a result of the finite box size constraining the solution, since for small $\phi$ a very large box is required to avoid convective plumes artificially leaving one side of the box and entering on the other.

$\delta T_{rms}$ in the simulations tends to slightly increase between $\phi=90^{\circ}$ and $\phi\approx40^{\circ}$. This trend is reasonably well captured by the multi-mode theory but not by the single-mode theory which decreases monotonically with decreasing $\phi$. Again, at the smallest $\phi$, the behaviour of $\delta T_{rms}$ is not well described by the theoretical predictions; the reasons for this have been touched upon above and will be pursued further in section \ref{Meanflows}.

Comparison between simulation and theoretical $\hat{k}$ is given in Fig.~\ref{fig_fixOm10_other} (e) and (f). The theoretical prediction from single-mode RMLT given by equation (\ref{S793}) does not require additional normalisation. The multi-mode expression for $\hat{k}$ given by equation (\ref{khat}) is also exact and should not require additional normalisation. However, we find that some normalisation is required for the latter to agree with the data. This is probably a result of the subtle differences between the definitions in (\ref{khat}) and (\ref{khatsims}) (i.e., $F_\lambda$ represents the heat flux in a logarithmic interval in $k$-space, whereas $\hat{u}_{z,k}(k_x,k_y,t)\hat{T}^*_k(k_x,k_y,t)$ represents the heat flux in a unit interval in $k_x$ and $k_y$). Here the normalisation is such that the single- and multi-mode theories agree at the pole. 
In this case, there is little difference between the single and multi-mode predictions and both match the trend of the simulation data well; decreasing with decreasing $\phi$. 
Where the multi-mode theory does supercede the single mode theory is in describing the asymmetry in $k_x$ and $k_y$ -- as shown most clearly in Fig.~\ref{figf6}.

\subsection{Wavenumber scaling: MLT vs Linear Onset}\label{wavenumbersec}
Rotation affects both the vigour of convective flows and their spatial structure: as Coriolis forces become stronger, the convective eddies tend to align with the rotation axis, and to narrow in horizontal extent. (See, for example, discussion in \citealp{Gilman1975}.)  These effects manifest as changes in the typical wavenumber of the convection. In S79's single-mode RMLT, the dominant wavenumber (that is, the mode that transports the most heat) is expected to increase with rotation rate, scaling as $\Omega_z^{3/5}$ (see Appendix \ref{AppendixSingleMode}). This differs from the rotation-rate scaling of the most unstable mode at convective onset, obtained from linear theory\footnote{This strictly applies for fixed temperature boundary conditions on both boundaries in $z$, rather than one boundary with fixed flux and the other with fixed temperature, but we expect this difference to be unimportant even to the numerical prefactor.} \citep{Chandra1961},
\begin{eqnarray}\label{konset}
k_{\rm{onset}}=k_{x,{\rm{onset}}}=\left(\frac{\pi^2}{2}\right)^{\frac{1}{6}}Ek^{-\frac{1}{3}},
\end{eqnarray}
with $Ek$ as given in (\ref{tradnondim}) and where $k_{y,{\rm onset}}=0$ when $\phi\ne 90^{\circ}$.
Many previous studies have argued that the wavenumber variation in rotating convection simulations is at least approximately in accord with this linear scaling, even well into the nonlinear regime (e.g.,~\citealt{Tilgner09,Stellmach2014}; see also \citealt{Guervilly2019}). Crucially, the linear scaling depends on viscosity even at arbitrarily low $\nu$, whereas the RMLT prediction does not. When extrapolated to astrophysical or geophysical regimes, the two scalings can thus give very different predictions for the typical size of convective eddies, so clarifying which one applies, and in what regimes, is of considerable interest.

\begin{figure}
\begin{center}
\includegraphics[scale=1,trim = {0mm 0mm 0mm 0mm}, clip ]{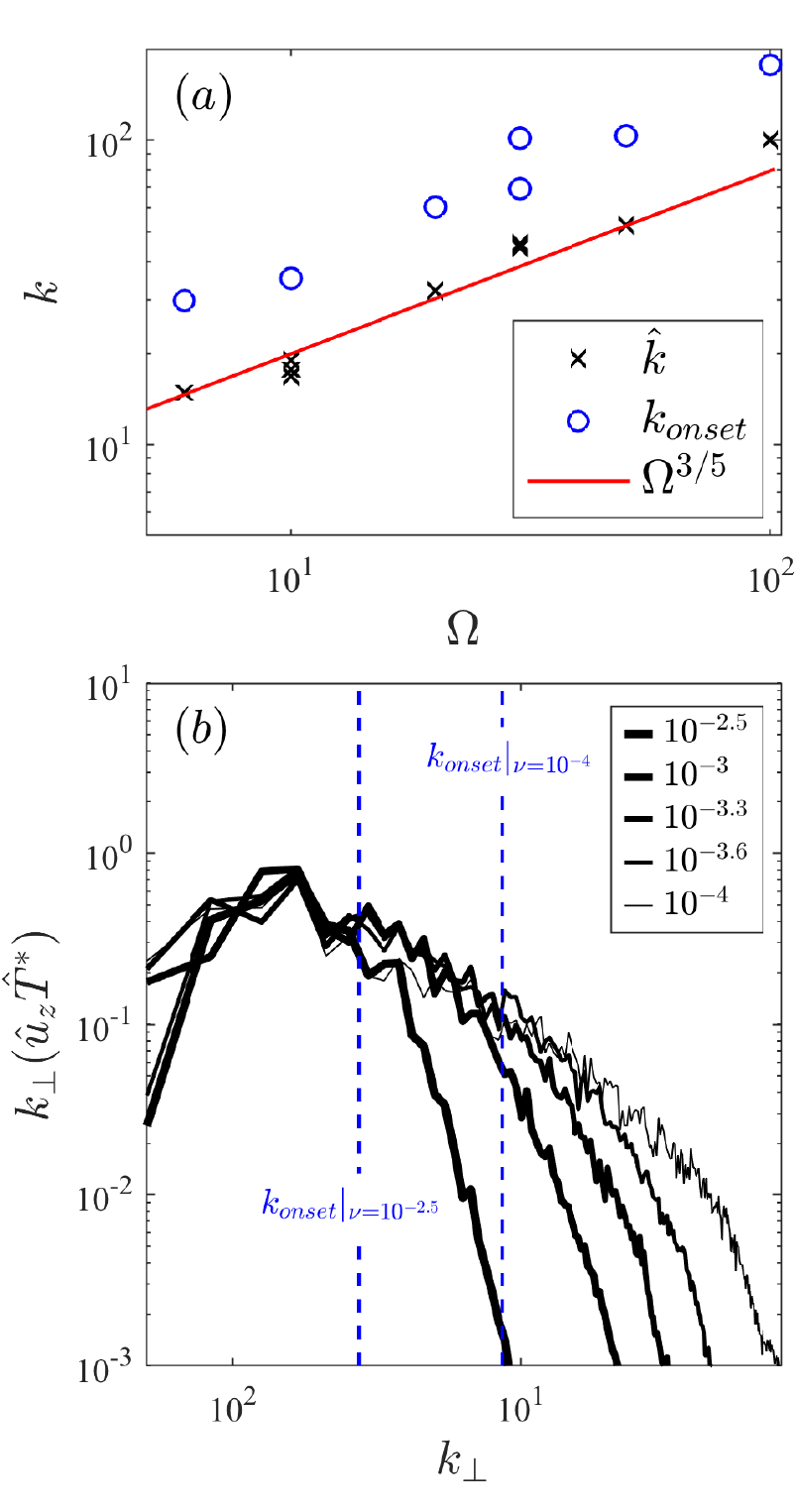}
\caption{(a) Variation of $\hat k$ (black crosses) and $k_{\rm{onset}}$ (blue circles) with $\Omega$ for a series of simulations with $\phi=60^{\circ}$. The red line shows the scaling of $\hat k$ expected from RMLT. (b) Heat flux spectra for five different $\nu$ at fixed $\Omega=10$, $\phi=45^{\circ}$, $L_x=L_y=1.5$ (black solid lines; the thicker the line, the larger $\nu$). The vertical lines show the value of $k_{\rm{onset}}$ for the largest (left) and smallest (right) viscosities considered here; $k_{\rm{onset}}$ varies with $\nu$ whereas the peaks of the spectra do not.}\label{fig_khatscaling}
\end{center}
\end{figure}

We therefore turn in Fig.~\ref{fig_khatscaling} to an assessment of the horizontal length scales present in our simulations, for simulations at varying rotation rates and viscosities. Fig.~\ref{fig_khatscaling}(a) shows how the dominant horizontal wavenumber, $\hat k$ (defined in (\ref{khatsims})) in a series of simulations at $\phi=60^{\circ}$ scales with rotation rate. 
For comparison, we have plotted $k_{\rm{onset}}$ as calculated by equation (\ref{konset}) above at each rotation rate, and have overplotted the RMLT theoretical scaling. It is clear, on the one hand, that the data are consistent with the RMLT prediction, which captures the broad trend with rotation rate (with $\hat{k}$ scaling roughly as $\Omega_z^{3/5}$).
However, the same data are also approximately in accord with the $Ek^{-1/3}$ scaling of linear theory: this is indicated by the $k_{\rm{onset}}$ points, which lie nearly parallel to the $\hat k$ values from the simulations. On the basis of this data alone, we thus cannot readily determine whether the RMLT wavenumber scaling (which does not depend on viscosity) or the linear onset one (which does) is the most appropriate. This is largely a consequence of numerical limitations: at each $\Omega$, we have been able to explore only a fairly narrow range of viscosities, so there is little evident distinction between a $\nu^{-1/3}$ scaling and a $\nu$-independent one. However, the quantitative predictions of the onset wavenumber are consistently larger than the numerical data.

A more detailed analysis, though, lends support to the view that the RMLT wavenumber scaling more accurately describes our simulations. We turn in Fig.~\ref{fig_khatscaling}(b) to an analysis of the heat flux power spectrum 
in a series of simulations at the same rotation rate ($\Omega = 10$) and latitude but varying $\nu$. If the linear onset scaling were appropriate, the "peak" of the power spectrum would be expected to move to the right (i.e., towards higher wavenumbers) as $\nu$ is decreased, since at fixed $\Omega$ this decreases $Ek$, as indicated by the dashed blue lines -- but this is not what is observed.  The power spectrum does change as $\nu$ is varied, but the largest changes are confined to high-wavenumber modes where the effects of viscosity are felt most keenly.  The peak of the spectrum remains largely unchanged.  (In general, the behaviour of the low-wavenumber modes is not utterly independent of those at high wavenumbers, of course; see, for example, discussions in \citet{FeatherstoneHindman2016}. However, this linkage need not change the peak wavenumber of the spectrum, or indeed the power of modes near that peak.)  We conclude that the rotating MLT scaling -- rather than the viscosity-dependent prediction of linear theory -- is the most appropriate one in explaining the dominant wavenumber of the convection in our simulations. 
Further, if the dominant wavenumber was determined by viscosity, as in the linear onset scaling, this would lead to different predictions for RMLT for other quantities such as $d\langle T\rangle/dz$. The internal self-consistency of RMLT thus demands that the dominant wavenumber follows the diffusion-free prediction.

\subsection{Connection to previous studies}\label{sec:connection}
\begin{figure*}
\begin{center}
\includegraphics[scale=1,trim = {0mm 0mm 0mm 0mm}, clip ]{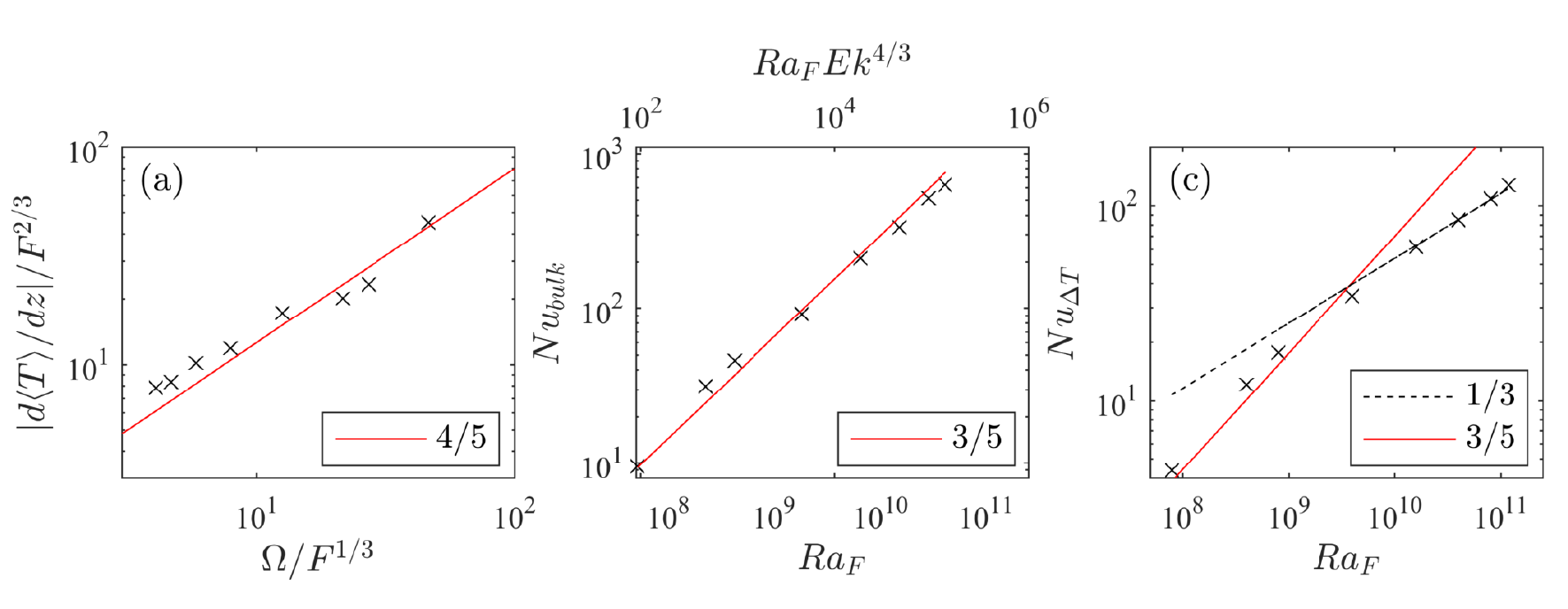}
\caption{The crosses represent a set of simulations using setup A in which $\phi=45^{\circ}$, $\nu=10^{-3.3}$, $\Omega=10$, and $F$ varies from .01 to 15; equivalently, $Ek=3.54\times 10^{-5}$ for a range of $Ra_F$. In (a) the bulk temperature gradient (measured over the middle one-third of the domain) is shown against the rotation rate, both expressed in "flux units". The red line represents the scaling predicted by RMLT. In (b), the same data is displayed in terms of a bulk Nusselt number against either $Ra_F$ (bottom axis) or $Ra_FEk^{4/3}$ (top axis). Again, the red line represents the scaling expected from RMLT. (c) shows a Nusselt number measured over the whole depth against $Ra_F$; the red line represents the scaling expected from RMLT and the black dashed line, the scaling as predicted by MLT without rotation.}\label{figNuRaF}
\end{center}
\end{figure*}

In prior sections, we chose "flux units" in which $F=1$ (in addition to setting $H=1$). By contrast, most previous studies have chosen to express their results in terms of the dimensionless quantities $Ra$, $Ek$, and $Nu$. To connect the two approaches, and to examine how our results compare to those in less-rapidly and more-rapidly rotating regimes, in Figure \ref{figNuRaF} we have briefly assessed the heat transport for a complementary set of simulations at $\phi=45^{\circ}$, in which we fix $\nu=10^{-3.3}$ and $\Omega=10$, and vary $F$ from $.01$ to $15$. We also set $H=1$ as before. Along this path in parameter space, $Ek=3.54\times 10^{-5}$, so that variations in $F$ correspond directly to changes in the supercriticality of the convection. (Note that changing $F$ is equivalent to keeping $F=1$ and changing $\Omega$ and $\nu$ such that their ratio is fixed.)

In Figure \ref{figNuRaF} (a), we show the results in flux units (see Table 1), plotting the temperature gradient normalised by $F^{2/3}$ against rotation rate normalised by $F^{1/3}$. This panel is essentially equivalent to Figure \ref{fig_fixphi}(c), except for the extension to lower values of $\Omega/F^{1/3}$, and for the slightly different path in parameters space taken here.  We see that the RMLT prediction (red line) begins to fail at small $\Omega$, as may be expected (see also Paper I).

In (b), we re-plot these results in terms of more traditional nondimensional measures of buoyancy driving and of heat transport, namely $Ra_F$ and $Nu_{bulk}=F/(\nu|d\langle T\rangle/dz|)$, (where $|d\langle T\rangle/dz|$ is the bulk temperature gradient as calculated in §3.1). Note that variations in $Ra_F$ are equivalent to variations in the viscosity in flux units (i.e., $Ra_F = ((\nu/(F^{1/3}H^{4/3}))^{-3}$).  The RMLT prediction from panel (a) becomes $Nu={\rm const\,} Ra_F^{3/5}$ and is shown by the red line.  Clearly, this scaling law holds for our simulations with $\phi=45^{\circ}$ over several decades in convective supercriticality. For ease of comparison with prior results, we have also shown the $x$-axis scaled by $Ek^{4/3}$ (see top axis of panel (b)).
As discussed in §1, the RMLT prediction in the rapidly rotating limit is, in this nondimensional view, equivalent to the relation $Nu \propto (Ra_F Ek^{4/3})^{3/5}$, derived in a very different manner by \citet{Julien2012} (extended here to the tilted case) and explored by many other authors.  Since $Ek$ is constant for these simulations, the two versions of the $x$-axis differ only by a constant factor (and are in turn a constant multiple of the convective supercriticality).

Instead of $Nu_{bulk}$, the heat transport in convection simulations is
often characterised by a Nusselt number defined over the full depth of the domain, i.e., $Nu_{\Delta T}=FL_z/(\nu\Delta T)$.  For comparison to such work, Figure \ref{figNuRaF} (c) replots panel (b), but using $Nu_{\Delta T}$. There appears to be a new regime at high $Ra_F$ (i.e., low $\Omega/F^{1/3}$, or high Rossby number), with a new power-law.  The slope of the power law in this regime is consistent with previously-proposed ``non-rotating'' scaling relations ($Nu \propto
Ra_F^{1/3}$, or $Nu \propto Ra_F^{1/4}$); we have overplotted the $Nu \propto Ra_F^{1/3}$ scaling for comparison. Note that this is equivalent to the diffusion-free scaling of non-rotating MLT.  However, we argue that this measure provides a somewhat misleading view of the convective transport.  Naively, one might expect that $Nu_{\Delta T}$ is an adequate proxy for $Nu_{bulk}$, because they differ from each other by the factor $dT/dz / (\Delta T/L_z)$, which might be expected to be around unity. But if a simulation possesses thin thermal boundary layers at the top and bottom of the computational domain, across which the temperature falls considerably, this factor can become very
small.  The apparent break in Figure \ref{figNuRaF}(c) arises because these boundary layers ``throttle'' the overall transport at slow rotation rates; at higher rotation rates, the bulk transport dominates, and so measures of $Nu_{bulk}$ or $Nu_{\Delta T}$ give comparable results. See also Paper 1 for further discussion of the role of boundary layers.

\section{Effects of zonal flows on heat transport}\label{Meanflows}
In section \ref{compsim} we demonstrated that multi-mode RMLT does a reasonably good job in describing the behaviour of the bulk properties of convection, however, some discrepancies remain.
We believe there are two main reasons for these discrepancies. First, as we have already discussed, the periodic boundaries in $y$ are likely to strongly constrain the flow when $\cot\phi\gtrsim L_y/L_z$, which is always satisfied when $\phi$ approaches $0^{\circ}$. This means that simulations very close to the equator cannot reliably test RMLT. Second, large-scale flows are generated by the convection as alluded to in section \ref{sims}, and these flows are not accounted for in the theory presented in section \ref{MLT}. In figures \ref{fig_fixOm1030} and \ref{fig_fixOm10_other}, we highlighted the cases with strongest zonal flows (grey circles) and demonstrated that these have the largest discrepancies from the theoretical predictions. In this section we briefly describe the large-scale flows, and investigate their effects on heat transport further.

\subsection{Generation of large-scale flows: effects of varying the horizontal box sizes}\label{LSflows}
\begin{figure*}
\includegraphics[scale=0.95,trim = {0mm 0mm 0mm 0mm}, clip ]{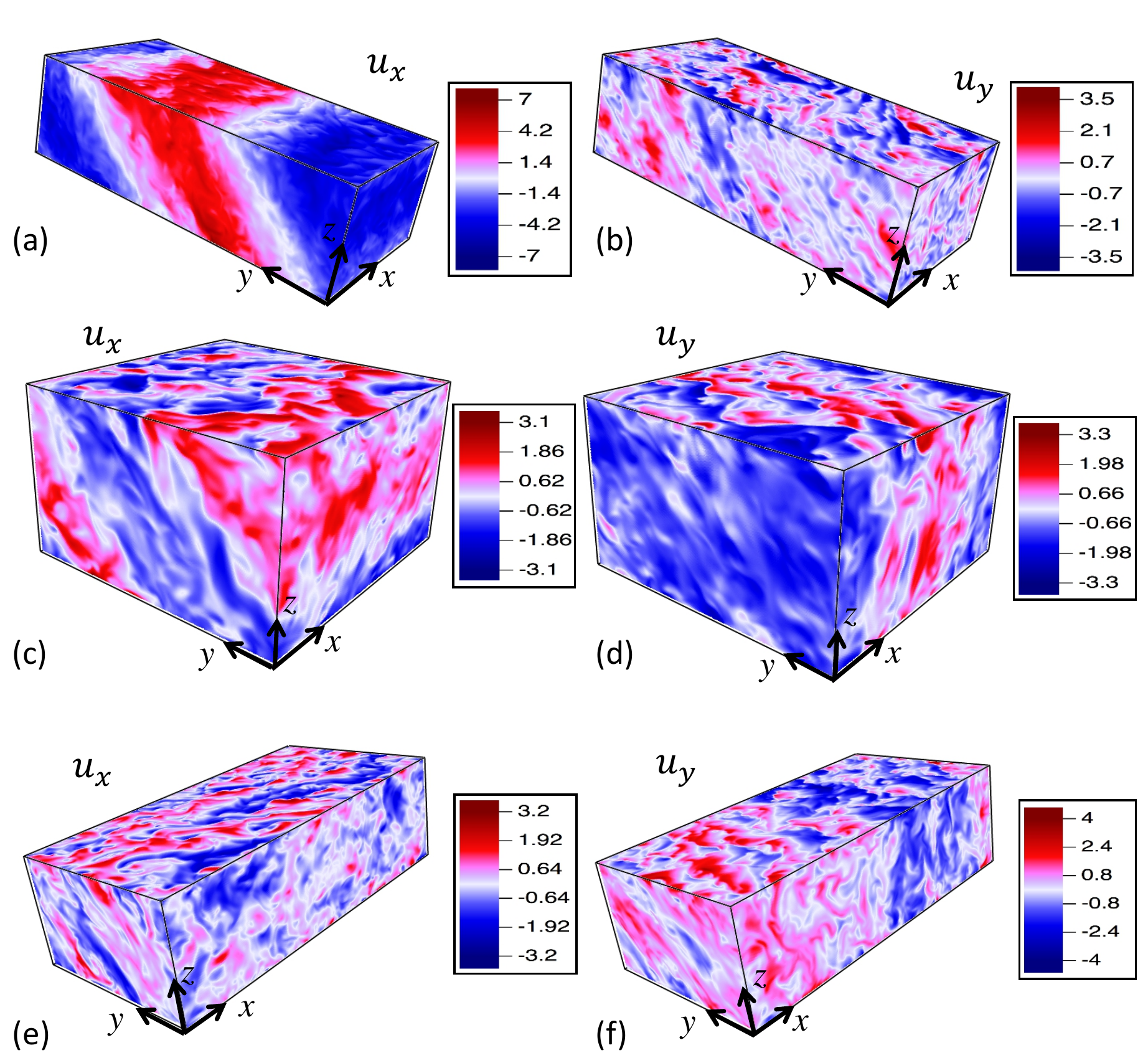}
\caption{Snapshots of the horizontal velocity components for cases 10A45e (a-b), 10A45c (c-d) and 10A45h (e-f). (a), (c) and (e) are zonal velocities ($u_x$) and (b), (d) and (f) are meridional velocities ($u_y$). In all cases $\Omega=10$ and $\phi=45^{\circ}$. In (a) and (b), $L_x=2, L_y=4$, in (c) and (d), $L_x=2, L_y=2$ and in (e) and (f), $L_x=4, L_y=2$. When $L_y>L_x$ strong zonal jets are visible which are aligned in the $x$-direction, for $L_y\sim L_x$ both coherent meridional and zonal jets are visible and for $L_x>L_y$ meridional jets aligned with the $y$-axis are visible.}\label{figf8}
\end{figure*}

In our simulations we observe significant large-scale zonal and meridional jets. 
We find that the strength and direction of these flows is strongly dependent on the horizontal box sizes $L_x$ and $L_y$ (see Fig.~\ref{figf8}).  This dependence on box size highlights the somewhat artificial nature of the mean flows realised in our simulations; however, it means that by varying box sizes at fixed rotation rate and latitude, we can systematically excite flows of different strength, and explore how these influence the heat transport, in a manner that would be difficult or impossible in other setups.  (In reality, and in some global calculations, zonal flows arise naturally as a consequence of convective angular momentum transport, and their strength cannot easily be "dialed in".)  In this sub-section, we briefly describe the strength of the flows and how this depends on the aspect ratios of our simulated system; in the next, we discuss how the flows affect the heat transport.

In Fig.~\ref{figf8}, we sample the zonal and meridional flows realised in a few different calculations at $\Omega=10$ and $\phi=45^{\circ}$, but with varying box sizes $L_x$ and $L_y$ in the two horizontal dimensions.  
In general, we find that if $L_y>L_x$, meridional flows ($u_y$) are suppressed and zonal flows ($u_x$) are enhanced, and if $L_x>L_y$ zonal flows are suppressed and meridional flows enhanced. When $L_x=L_y$, strong jets are observed in both directions, which may correspond with a large scale vortex. The strongest jets are therefore always aligned parallel to the shortest side, as also observed by \cite{Guervilly2017} in their simulations at the pole.

The example zonal jets in Fig.~\ref{figf8} are quasi-geostrophic flows that do not vary along $\Omega$, so they are tilted in the $(y,z)$-plane (similar zonal jets have also been obtained recently by \citealt{Novi2019}). The zonal (and meridional) jets exhibit a preferred wavelength in $y$ ($x$), which is much larger than the typical wavelengths of the dominant convective lengthscales. 

We have observed that if $L_y$ (or $L_x$) is increased for a fixed $L_x$ ($L_y$), the zonal flow (meridional flow) strength tends to become approximately independent of $L_y$ (or $L_x$) once we contain at least one full wavelength of this structure in $y$ ($x$), however the strength of these flows has not saturated in time in all of our simulations. In addition, the strength of these flows does depend on the horizontal aspect ratio $L_x/L_y$.

In general, these flows arise as a consequence of the organised Reynolds stresses within the rotating convection.   We defer a detailed analysis of the generation and saturation of the flows to other work, 
but note that the equilibrated amplitude of the flows may be expected to depend both on the convective Reynolds stresses and viscosity, and also on the presence or absence of "parasitic" instabilities that can sap the energy of these flows. 

Each of the simulations listed in Table~\ref{Table1} have been run for a minimum time interval of 30 time-units (and up to 200 time-units) once a turbulent convective quasi-steady state has been reached. The run time of our simulations was found to be sufficient to obtain adequately converged statistics for bulk properties such as the mean temperature gradient (i.e., those discussed in sections \ref{sims} and \ref{MLT}). 
However, the development of the large scale flows can take many convective turnover times (particularly when $L_x=L_y$) and so some of our simulations could still be undergoing longer-term behaviour such as jet merging on a viscous time-scale (e.g. \citealt{Guervilly2017}), but capturing these effects is not our primary focus.
Nevertheless, the example flows shown in Fig.~\ref{figf8} are representative of those that we observe in our simulations, and similar results are found  with both setups.

\begin{figure}
\includegraphics[scale=1,trim = {0mm 0mm 0mm 0mm}, clip]{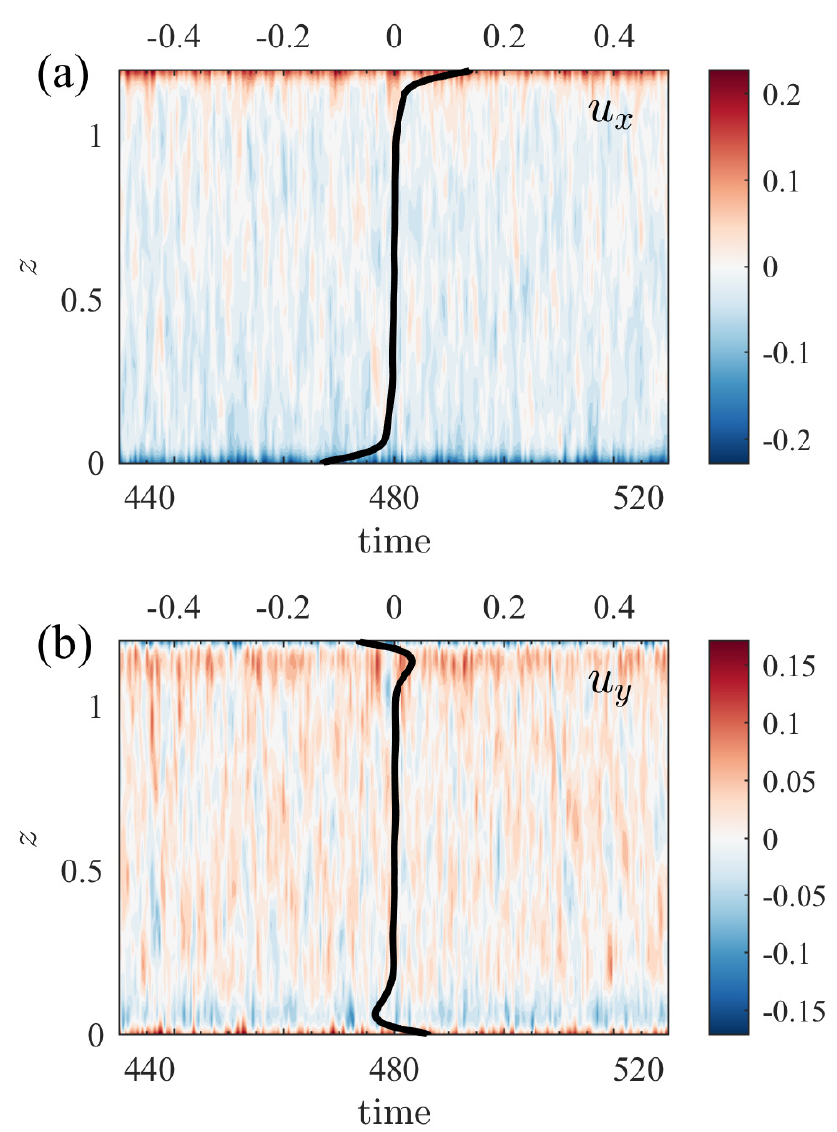}
\caption{Horizontally-averaged $x$ and $y$ components of $u_x$ (a) and $u_y$ (b) as a function of $z$ and time (bottom $x$ axis). The corresponding time-averaged profiles as a function of depth are given by the overlying thick black lines and their values are given by the top $x$ axis. In this case the parameters are taken to be the same as those used in Fig.~\ref{figf8} (a), (b) (case 10A45e in Table~\ref{Table1}). The time-averaged mean flows are small in the interior but are more significant in the boundary layers; this is typical of all our simulations.}\label{figf7}
\end{figure}

Unlike in \citet{HS1983}, we observe no significant mean (i.e., $x$ and $y$-averaged) flows in our simulations except very close to the boundaries. This can be seen in Fig.~\ref{figf7} in an example simulation, where the contours show $\langle u_x \rangle_{xy}$ (a) and $\langle u_y\rangle_{xy}$ (b) as a function of $z$ and time (here averages of a quantity over both $x$ and $y$ are represented by $\langle \cdot\rangle_{xy}$). The over-lying solid black lines show the time average of these quantities as a function of $z$. Clearly, whilst at any one time there can be non-trivial horizontally-averaged flows in the bulk, these are oscillatory and cancel out on space and time-averaging to leave only a very weak mean flow. Close to the boundaries where the flows are more laminar there are systematic flows similar to those seen in \citet{HS1983} except that they observe significant mean flows throughout the entire domain (not just close to the boundaries). We attribute this difference to the more rapidly rotating, turbulent regime in which our simulations lie (similar effects on the mean flow of rotation and turbulence have been reported by \citet{Brummell1996} in compressible convection, for example). Indeed, we do find more systematic mean flows if we increase $\nu$ (equivalently $Ek$). The picture in Fig.~\ref{figf7} is typical of all our rapidly rotating simulations, and very similar results are also found using both setups. This indicates that rapidly rotating turbulent convection in the setups considered here does not generate significant, persistent $x$ and $y$-averaged mean flows, except close to boundaries. (Note that periodic boundary conditions prevent the convection from generating mean horizontal temperature gradients in $x$ and $y$.)

\subsection{Bulk properties as a function of large-scale flow strength}\label{results2}

To investigate the effect of the large-scale flows on the bulk convective properties we first characterise the strength of the flow by defining the RMS value of each horizontal component;
\begin{equation}\label{xirms}
\xi_{rms}=\sqrt{\langle \xi^2 \rangle_{xyz}}
\end{equation}
where $\xi$ is taken to be $u_x$ or $u_y$ and the subscript on the angled brackets denotes the coordinates that are averaged over. In addition, we consider the following two measures
\begin{equation}\label{zonal}
\langle u_x \rangle_{x,{rms}}=\sqrt{\langle (\langle u_x \rangle_x)^2 \rangle_{yz}},
\end{equation}
\begin{equation}\label{merid}
\langle u_y \rangle_{y,{rms}}=\sqrt{\langle (\langle u_y \rangle_y)^2 \rangle_{xz}}.
\end{equation}
Since the zonal ($u_x$) jets extend over the entire domain in $x$ and alternate in the $y$-direction, and similarly, the meridional ($u_y$) jets extend over the entire domain in $y$- and alternate in $x$, the quantities $\sqrt{\langle (\langle u_x \rangle_y)^2 \rangle_{xz}}$ and $\sqrt{\langle (\langle u_y \rangle_x)^2 \rangle_{yz}}$ are small and so we do not consider these further.

\begin{figure*}
\includegraphics[scale=1,trim = {0mm 0mm 0mm 0mm}, clip ]{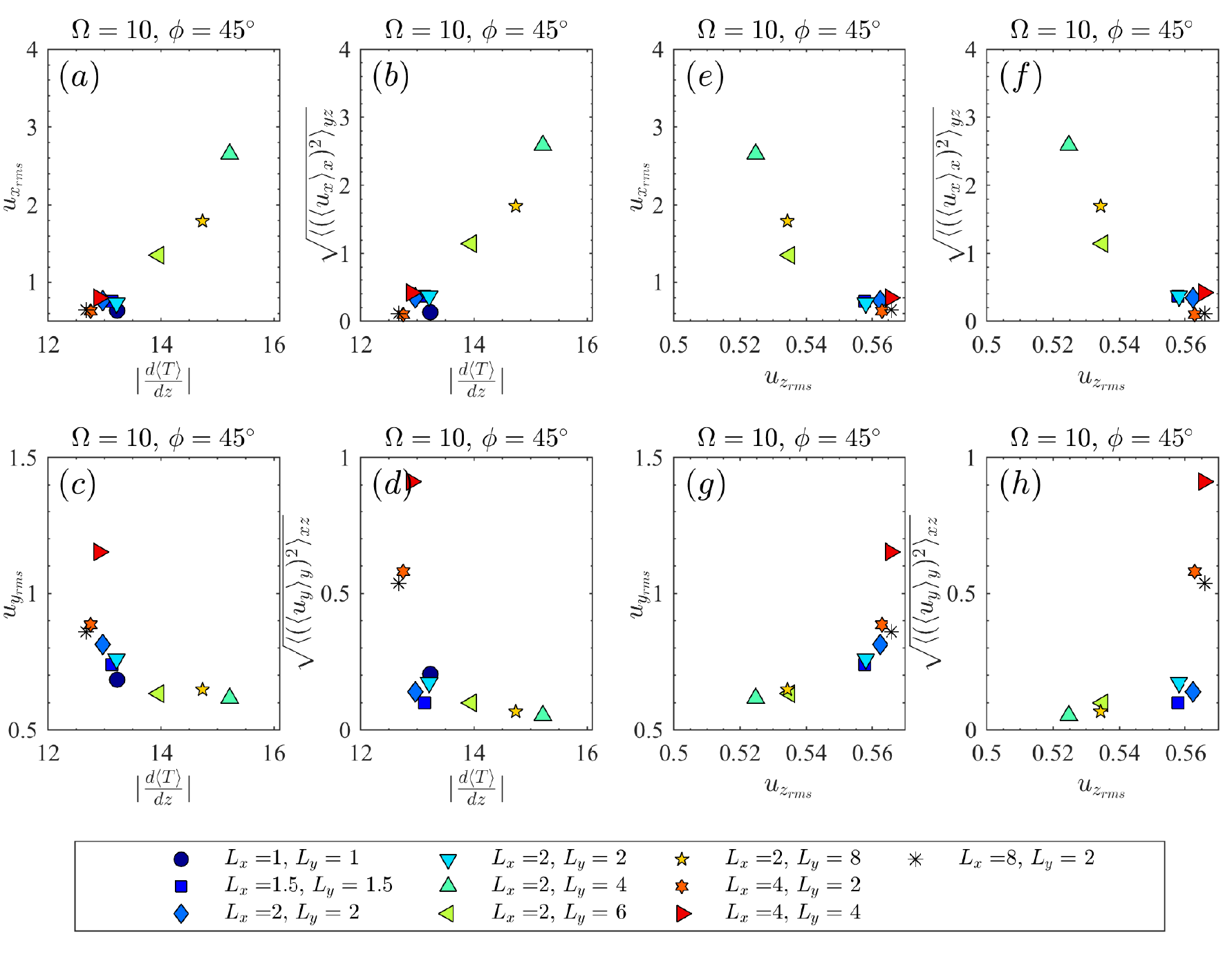}
\caption{Measure of the large-scale flow strength as a function of $|\frac{d\langle T\rangle}{dz}|$ (a-d) and as a function of $u_{z,rms}$ (e-h). In all cases $\Omega=10$ and $\phi=45^{\circ}$ but the horizontal box size and aspect ratio are varied which leads to different flows. There is a strong correlation between zonal flow strength as measured using (\ref{xirms}) or (\ref{zonal}) and $|\frac{d\langle T\rangle}{dz}|$ and $u_{z,{rms}}$; stronger flows lead to higher $|\frac{d\langle T\rangle}{dz}|$ and lower $u_{z,{rms}}$. The correlation between the meridional flow strength is much weaker though the strongest flows have the opposite effect to the zonal flows and correspond to lower $|\frac{d\langle T\rangle}{dz}|$ and higher $u_{z,{rms}}$. The data used here is from cases 10A45(a-k) in Table~\ref{Table1}.}\label{figf9}
\end{figure*}

In Fig.~\ref{figf9} we plot the measures of the large-scale flow given by (\ref{xirms})-(\ref{merid}) against the midplane temperature gradient and rms vertical velocity. There is a strong correlation between the zonal flow strength (either measured by $u_{x_{rms}}$ or $\langle u_x \rangle_{x_{rms}}$) and the average bulk temperature gradient with stronger flows corresponding to larger temperature gradients in the bulk. There is also a strong correlation between the vertical velocities and the zonal flow strength, whereby the stronger flows correspond to slower vertical velocities.
We interpret that the large-scale zonal flow inhibits the convection, leading to slower vertical velocities, and a larger temperature gradient to carry the same heat flux through the domain. The inhibiting effects of zonal flows on convection have been observed in other systems \citep[see e.g.,][]{Teedetal2012, Goluskinetal2014}, and proposed relations between shear and heat transport also figure prominently in the theoretical models of \citet{Balbusetal2009}.

The correlation between the meridional flow strength and the temperature gradient (and vertical velocity) does not exhibit the same behaviour. In fact the largest temperature gradients and smallest vertical velocities occur at the smallest values of $u_{y_{rms}}$ or $\langle u_y \rangle_{y_{rms}}$ (where the zonal flows are dominant). This indicates that it is the zonal flows that have the most important effects on inhibiting heat transport.

\begin{figure}
\includegraphics[scale=0.9,trim = {0mm 0mm 0mm 0mm}, clip ]{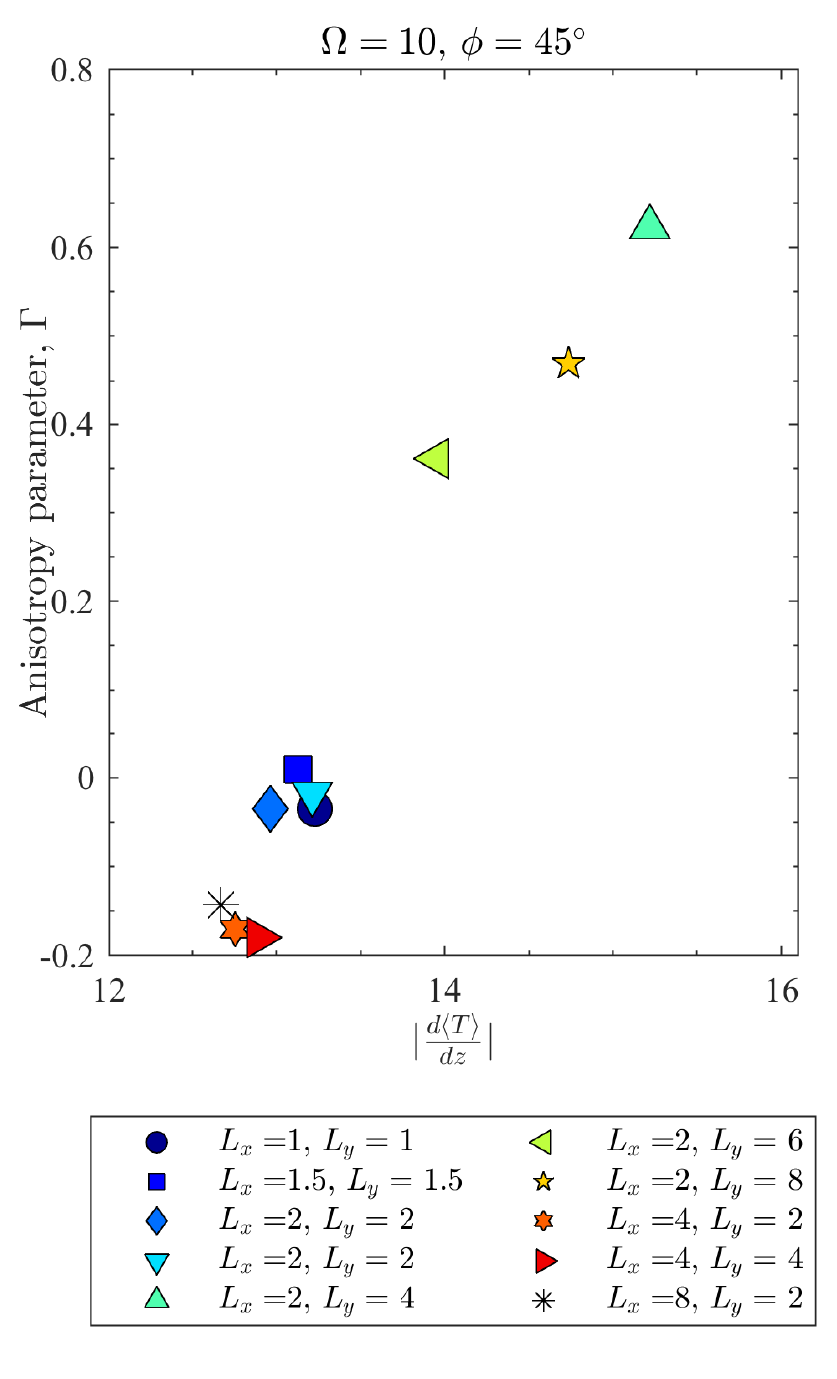}
\caption{Measure of the anisotropy in the large-scale flows as quantified by $\Gamma$ (defined in (\ref{Gamma}) as a function of $|\frac{d\langle T\rangle}{dz}|$. In all cases $\Omega=10$ and $\phi=45^{\circ}$ but the horizontal box size and aspect ratio are varied which leads to different flows. Cases with stronger zonal flow ($\Gamma>0$) impact the heat transport the most and correspond to a larger $|\frac{d\langle T\rangle}{dz}|$. The data used here is from cases 10A45(a-k) in Table~\ref{Table1}}\label{figf10}
\end{figure}

Fig.~\ref{figf9} shows that the strongest flows tend to be zonal (in the $x$-direction); we can investigate the anisotropy in the two horizontal directions with the scalar $\Gamma$ defined as follows:
\begin{equation}\label{Gamma}
\Gamma=\frac{u_{x_{rms}}-u_{y_{rms}}}{u_{x_{rms}}+u_{y_{rms}}}.
\end{equation}
A positive value of $\Gamma$ corresponds to cases where the zonal flow is stronger than the meridional flow, whereas, a negative value of $\Gamma$ corresponds to cases where the zonal flow is weaker than the meridional flow, and values close to zero indicate the meridional and zonal flows are roughly of equal strength.
$\Gamma$ is plotted against $|d\langle T\rangle/dz|$ in Fig.~\ref{figf10}. 
The magnitude of $\Gamma$ is larger for cases in which the sign of $\Gamma$ is positive.
This plot highlights that for values of $\Gamma$ that are large and positive, the temperature gradient is increased (convection is less efficient). For values of $\Gamma$ close to zero the temperature gradient remains roughly the same. For negative $\Gamma$ the temperature gradient is reduced from the cases with $\Gamma=0$. This again demonstrates that in these cases the large-scale (meridional) flows do not increase the mid-layer temperature gradient in the same way as the large-scale zonal flows. Note the criterion for highlighting which simulations in figures \ref{fig_fixOm1030} and \ref{fig_fixOm10_other} had significant zonal flow was $\Gamma>0.1$.

Our conclusion that the heat transport is strongly affected by the zonal flow qualitatively agrees with the findings of \citet{Guervilly2017} who considered local box simulations at the poles with $L_y\geq L_x$. They find that heat transport is reduced in cases with a large-scale flow and argue that increases in the local rotation rate, rather than shearing of the convection by zonal flows, cause the inhibition of heat transport.
Likewise, some spherical shell models have shown the effect of large-scale flows on heat transport may be quite complex \citep[see e.g.,][]{Yadavetal2015, Raynaudetal2018}. We defer a more detailed analysis of the relationship between heat transport and the zonal flows in our simulations to future work. 

\section{Discussion \& Conclusions}\label{conc}

We have examined the effect of rotation on the temperature gradient established by convection in a layer in which rotation and gravity are misaligned.  Our 3D simulations, though highly idealised representations of a small portion of a star or planet at a given latitude, have yielded new constraints on how the convective heat transport is influenced by rotation, by changes in latitude, and by other physical and numerical effects (including zonal flows).

In particular, we have argued that many aspects of the convection -- including, crucially, the temperature gradient it establishes -- are (in cases with weak zonal flow)
well-described by "rotating mixing length theory" (RMLT).  Our version of RMLT is fundamentally akin to that developed by \citet{Stevenson1979} and explored in Paper I: at its core, it assumes (as they did) that the equilibrated amplitude of each mode is set by equating its \emph{linear} growth rate (which can be calculated analytically) to its nonlinear cascade rate, and also employs the linearised equations to link the temperature fluctuations at a given point to the background temperature gradient and the velocity field.  The theory of S79 further simplified matters by assuming that the flow was dominated by the mode that transports the most heat, allowing for an analytical solution for $d\langle T\rangle /dz$ (and other quantities) as a function of rotation rate.  Our simulations confirm that the overall scaling implied by this single-mode theory, which predicts that $d\langle T\rangle/dz \propto \Omega^{4/5}$ in the rapidly rotating limit, describes our data well at all latitudes (see, e.g., Fig.~\ref{fig_fixphi}), extending prior work that studied only the polar region (Paper I) and complementing results on heat transport in rotating spherical shells \citep[e.g.,][]{Gastine2016}.  

However, our simulations have also revealed that at latitudes far from the poles, the single-mode theory fails to capture the horizontal anisotropy of the heat transport, which in turn leads to less-accurate estimates of $d\langle T\rangle/dz$ and other quantities.  We have thus turned to a multi-mode theory, described in Section \ref{MLT}, which provides predictions for the 
temperature gradient that, in cases with weak zonal flows, gives better agreement with our simulations with varying latitudes -- see, for example, Fig.~\ref{fig_fixOm1030}.

The theory also predicts other aspects of the convective flow, including the spectrum as a function of wavenumber and typical velocity and temperature fluctuations.  Some of these predictions (e.g., for the heat flux spectrum — see Fig. 8) are likewise in reasonably good accord with the results of our simulations, but some others (e.g., for the fluctuating temperature field as a function of latitude — see Fig. 9) do not agree especially well.  However, for all quantities, the largest disagreements between RMLT and the simulations tend to be found in cases with strong zonal flows, as discussed further below.

Although in extending the theory to include multiple modes one loses the ability to write down closed-form analytic expressions for the bulk properties as a function of rotation rate, the multi-mode theory is still quick and easy to compute (in particular, orders of magnitude cheaper than running full 3D simulations).  This suggests that some version of the theory, appropriately extended, may ultimately be suitable for inclusion in 1D evolutionary models of stars or planets, in place of current MLT models that do not include the effects of rotation at all. It must be reiterated, though, that some aspects of the convective dynamics may not be well-captured by any variant of MLT, including the multi-mode RMLT studied here---such as overshooting and upflow/downflow asymmetry.  More sophisticated models (including the closure-based theories of \citet{Xiongetal1997, Canuto2011}, or indeed full numerical simulations), may be required to
capture such effects.

Of course our simulations operate in parameter regimes very different from those in any real astrophysical object, and it is appropriate to examine the extent to which our results are influenced by this.  In Section \ref{supercrit}, we briefly assessed how our results depend on the artificially high viscosities in our calculations, finding that our models appear to be approaching a regime in which diffusion-free scalings provide an apt description. We also showed in section \ref{sec:connection} that our results hold over several decades of supercriticality.

We must caution, though, that many effects that are present in real stars or planets, but absent from the RMLT presented here, likely have a significant effect on the convective heat transport.  As an example, we point to the zonal flows (differential rotation) analyzed in Section \ref{Meanflows}.  In our simulations, the magnitude and character of these flows is dependent on the aspect ratio and size of the computational domain; this dependence, though artificial, in turn allowed us to examine how the heat transport changes as the zonal flow is altered, while keeping other factors (namely the rotation rate, diffusivities, and heat flux) constant.  We showed (see, e.g., Fig.~\ref{figf10}) that stronger zonal flows inhibit the convective transport, leading to larger temperature gradients and significant departures from the predictions of RMLT.  These results may ultimately constrain models of stellar convection in which the interplay between shear and heat transport is an essential element (e.g., \citealt{Balbusetal2009}).

Still other effects are entirely absent from our simulations but surely play a role in astrophysical objects.  Among these, the overall spherical geometry of stars or planets (and the $\beta$-effect of latitudinally-varying Coriolis parameter), their strong density stratification, and their ubiquitous magnetism are likely to be particularly significant.  By adopting the Boussinesq approximation, we have neglected the effects of compressibility and imposed a symmetry between upflows and downflows that does not exist in the full system, which may well have implications for the stratification that is established \citep[e.g.,][]{Korreetal2017, Kapylaetal2017}. Moreover, in strongly stratified systems, dissipative heating can be large resulting in convective fluxes that greatly exceed the luminosity \citep[see e.g.,][]{CurrieBrowning2017}. Furthermore, when stratification is present, the convective velocities will also generally vary with depth, so the influence of rotation on the dynamics may be depth-dependent as well \citep[see, e.g., discussions in][]{IrelandBrowning2018}. By using an f-plane model, we have considered only single latitudes in isolation from one another; in reality, the behaviour at different latitudes is coupled by large-scale flows.  We have not accounted for, or imposed, thermal variations in latitude, which are undoubtedly present in real stars and will in many cases lead to a thermal wind \citep[e.g.,][]{Rempel2005, Mieschetal2006}.  Finally, most stars or planets are affected by magnetism at some level, with this acting to hinder convective transport in some regimes and to aid it in others \citep[e.g.,][]{Chandra1961}; these effects are likewise not considered here.  Future work including these processes -- for example, 3D spherical calculations with strong stratification and magnetism -- will be necessary to examine how robust the conclusions drawn here really are.  As noted in \S 1, already-extant models along these lines \citep[e.g.,][]{Yadavetal2015,Gastine2016} show broadly good agreement with many of our results, albeit in a somewhat different parameter regime, but many issues remain to be explored.  

We conclude by reiterating that the simulations here, though highly idealised, nonetheless appear to capture some important facets of the interaction between convection, rotation, and shear -- processes which figure, alongside many others, in the interiors of virtually every star or planet.  The agreement between these simulations and the simplified RMLT presented here, though far from perfect, is remarkable given the simplicity of the theory and the ease with which it may be computed.  We hope in future work to explore how the many dynamical elements not captured here -- including magnetism and density stratification -- affect these results, and hence to assess their relevance for the evolution and structure of astrophysical objects.



\section*{Acknowledgements}
LKC \& MKB were supported by the European Research Council under ERC grant agreement No. 337705 (CHASM). LKC also acknowledges support from STFC Grant ST/R000891/1. AJB was supported by the Leverhulme Trust through the award of an Early Career Fellowship and by STFC Grants ST/R00059X/1 and ST/S000275/1. YL acknowledges NASA grant NNX14AD21G and NSF grant AST-1352369.
We also thank the anonymous referee for reviewing the manuscript.
The calculations for this paper were performed on the University of Exeter Supercomputers, Zen and Isca. The former was a DiRAC Facility jointly funded by STFC, the Large Facilities Capital Fund of BIS, and the University of Exeter; the latter is part of the University of Exeter High-Performance Computing (HPC) facility. 
Additional simulations were performed by the DiRAC Complexity system and the DiRAC Data Intensive service at Leicester, both operated by the University of Leicester IT Services, which forms part of the STFC DiRAC HPC Facility (www.dirac.ac.uk). This equipment is funded by BIS National E-Infrastructure capital grants ST/K000373/1 and ST/R002363/1 and STFC DiRAC Operations grants ST/K0003259/1 and ST/R001014/1. 
Simulations were also undertaken on ARC1, ARC2 \& ARC3, part of the High Performance Computing facilities at the University of Leeds, UK, including the DiRAC 1 Facility at Leeds jointly funded by STFC and the Large Facilities Capital Fund of BIS. This work also used the DiRAC Data Centric system at Durham University as part of the UK MHD Consortium, operated by the Institute for Computational Cosmology on behalf of the STFC DiRAC HPC Facility. This equipment was funded by a BIS National E-infrastructure capital grant ST/K00042X/1, STFC capital grant ST/K00087X/1, DiRAC Operations grant ST/K003267/1 and Durham University. 
DiRAC is part of the National e-Infrastructure. 
We are also grateful for prior access to PRACE for awarding us access to computational resources, namely Mare Nostrum based in Spain at the Barcelona Supercomputing Center, and Fermi and Marconi based at Cineca in Italy. 
The 3d renderings in figures \ref{plumes} and \ref{figf8} were created using VAPOR \citep{vapor}.





\begin{thebibliography}{}
\bibliographystyle{mnras}
\bibitem[Anders et al.(2019)]{Andersetal2019} Anders E.H., Manduca C.M., Brown B.P., Oishi J.S.,  Vasil, G.M., 2019, \apj, 872(2), p.138.
\bibitem[Arnett et al.(2009)]{Arnettetal2009} Arnett, D., Meakin, C. and Young, P.A., 2009.  \apj, 690(2), p.1715.
\bibitem[Arnett et al.(2019)]{Arnettetal2019} Arnett, W.D., Meakin, C., Hirschi, R., Cristini, A., Georgy, C., Campbell, S., Scott, L.J., Kaiser, E.A., Viallet, M.,  Moc√°k, M., 2019. ApJ, 882(1), p.18.
\bibitem[Aubert et al.(2001)]{Aub01} Aubert J., Brito D., Nataf H.-C., Cardin P., Masson, J.-P., 2001, Physics of the Earth and Planetary Interiors, 128, 51
\bibitem[Balbus et al.(2009)]{Balbusetal2009} Balbus S.A., Bonart J., Latter H.N., Weiss, N.O., 2009, Monthly Notices of the Royal Astronomical Society, 400(1), pp.176-182.
\bibitem[Baraffe et al.(2015)]{Baraffeetal2015} Baraffe I., Homeier D., Allard F., Chabrier G., 2015,  Astronomy \& Astrophysics, 577, p.A42
\bibitem[Barker et al.(2014)]{Barkeretal2014} Barker A. J., Dempsey A. M.  Lithwick, Y., 2014, \apj, 791, 13
\bibitem[B\"{o}hm-Vitense(1958)]{BV58} B\"{o}hm-Vitense E., 1958, Zeitschrift fr Astrophysik, 46, 108
\bibitem[Brandenburg(2016)]{Brandenburg2016} Brandenburg, A., 2016. \apj, 832(1), p.6.
\bibitem[Brummell et al.(1996)]{Brummell1996} Brummell N. H., Hurlburt N. E., Toomre J., 1996, ApJ, 473, 494f
\bibitem[Brummell et al.(1998)]{Brummell1998} Brummell N. H., Hurlburt N. E., Toomre J., 1998, ApJ, 493, 955
\bibitem[Brummell et al.(2002)]{Brummell2002} Brummell N. H., Clune T. L., Toomre J., 2002, ApJ, 570, 825
\bibitem[Brun \& Browning(2017)]{BrunBrowning2017} Brun A.~S.,  Browning M.~K., 2017, Living Reviews in Solar Physics, 14, 4
\bibitem[Burns et al.(2019)]{Dedalus2019} Burns K.~J., Vasil G.~M., Oishi J.~S., Lecoanet, D., Brown B.~P., 2019, arXiv: 1905.10388, submitted
\bibitem[Canuto(1996)]{Canuto1996} Canuto, V.~M., 1996 \apj, 467, p.385.
\bibitem[Canuto(2011)]{Canuto2011} Canuto, V.~M., 2011. Astronomy \& Astrophysics, 528, p.A76.
\bibitem[Cattaneo et al.(1991)]{Cattaneo1991} {Cattaneo} F., {Brummell} N.~H., {Toomre} J., {Malagoli} A. {Hurlburt} N.~E., 1991, ApJ, 370, 282
\bibitem[Chabrier et al.(2007)]{Chabrieretal2007} Chabrier G., Gallardo J., Baraffe I., 2007, Astronomy \& Astrophysics, 472(2), pp.L17-L20.
\bibitem[Chandrasekhar(1961)]{Chandra1961} Chandrasekhar S., Hydrodynamic and hydromagnetic stability, International Series of Monographs on Physics, Oxford: Clarendon, 1961
\bibitem[Chan(2007)]{Chan2007} Chan K. L., 2007, AN, 328, 10, 1059
\bibitem[Christensen et al.(2002)]{Christensen2002} Christensen U. R., 2002, JFM, 470, 115
\bibitem[Christensen \& Aubert(2006)]{ChrisAub06} Christensen U. R.,  Aubert J., 2006, Geophysical Journal
International, 166, 97
\bibitem[Clyne et al.(2007)]{vapor} Clyne J., Mininni P., Norton A., Rast, M., 2007 Mew Journal of Physics, 9(8), 301
\bibitem[Currie \& Tobias(2016)]{CurrieTobias2016} Currie L. K.,  Tobias, S.~M, 2016, Physics of Fluids, 28(1), p.017101.
\bibitem[Currie \& Browning(2017)]{CurrieBrowning2017} Currie L. K.,  Browning M.~K., 2017, \apjl, 845:L17, 7pp
\bibitem[Featherstone \& Hindman(2016)]{FeatherstoneHindman2016} Featherstone N.A., Hindman B.W., 2016, \apjl, 830(1), p.L15
\bibitem[Fischer et al.(2008)]{Nek5000} Fischer P. F., Lottes, J. W., Kerkemeier S. G., 2008, nek5000 Web page, http://nek5000.mcs.anl.gov
\bibitem[Flasar \& Gierasch(1978)]{FG78} Flasar F. M., Gierasch P. J., 1978, GApFD, 10, 175
\bibitem[Favier et al.(2014)]{Favier2014} Favier B., Silvers L. J. Proctor M. R. E., 2014, Physics of Fluids, 26, 9, 096605
\bibitem[Gastine et al.(2016)]{Gastine2016} Gastine T., Wicht J. Aubert J., 2016, JFM, 808, 690
\bibitem[Gillet \& Jones(2006)]{GilletJones2006} Gillet N.,  Jones C. A., 2006, JFM, 554, 343
\bibitem[Gilman(1975)]{Gilman1975} Gilman P.~A., 1975, Journal of Atmospheric Sciences, 32, 1331
\bibitem[Gilman(1977)]{Gilman1977} Gilman P.~A., 1977, Geophysical \& Astrophysical Fluid Dynamics, 8(1), pp.93-135.
\bibitem[Goluskin et al.(2014)]{Goluskinetal2014} Goluskin D., Johnston H., Flierl G.R., Spiegel E.A., 2014, Journal of Fluid Mechanics, 759, pp.360-385.
\bibitem[Gough \& Weiss(1976)]{GoughWeiss1976} Gough D.O.,  Weiss N.O., 1976. Monthly Notices of the Royal Astronomical Society, 176(3), pp.589-607.
\bibitem[Gough(1977a)]{Gough1977a} Gough, D.O., 1977, International Astronomical Union Colloquium, CUP
\bibitem[Gough(1977b)]{Gough1977b} Gough, D.O., 1977, \apj, 214, p.196
\bibitem[Guervilly \& Hughes(2017)]{Guervilly2017} Guervilly C., Hughes D. W., 2017, Physical Review Fluids, 2, 11, id.113503
\bibitem[Guervilly et al.(2014)]{Guervilly2014} Guervilly C., Hughes D. W., Jones C. A., 2014, JFM, 758, 407
\bibitem[Guervilly \& Cardin(2017)]{Guervilly2017a} Guervilly C., Cardin P., 2017, Geophys. J. Int., 211, 455
\bibitem[Guervilly et al.(2019)]{Guervilly2019} Guervilly, C., Cardin P., Schaeffer N., 2019, Nature, 570 (7761), 368
\bibitem[Hathaway et al.(1979)]{H79} Hathaway D. H., Toomre J., Gilman P. A., 1979, GApFD, 13, 289
\bibitem[Hathaway et al.(1980)]{H80} Hathaway D. H., Toomre J., Gilman P. A., 1980, GApFD, 15, 7
\bibitem[Hathaway \& Somerville(1983)]{HS1983} Hathaway D. H., Somerville R. C. J., 1983, JFM, 126, 75
\bibitem[Hotta et al.(2017)]{Hotta2016} {Hotta} H., {Rempel} M., {Yokoyama} T., 2016, Science, 351, 1427
\bibitem[Ingersoll \& Pollard(1982)]{IngPoll1982} Ingersoll A. P., Pollard D., 1982, Icarus, 52, 62
\bibitem[Ireland \& Browning(2018)]{IrelandBrowning2018} Ireland, L.G., Browning M.K., 2018, The Astrophysical Journal, 856(2), p.132.
\bibitem[Jermyn et al.(2018)]{Jermyn2018} Jermyn A. S., Lesaffre P., Tout C. A., Chitre S. M., 2018, MNRAS, 476, 646
\bibitem[Jones \& Kuzanyan(2009)]{JonesKuz09} Jones C. A., Kuzanyan K. M., 2009, Icarus, 204, 227
\bibitem[Jones et al.(2015)]{Jones2015} Jones C.A., 2015. Thermal and Compositional Convection in the Outer Core
In: Treatise on Geophysics 2nd Edition, volume 8, volume editor P. Olson, series editor G. Schubert. Elsevier.
\bibitem[Julien \& Knobloch(1998)]{JK98} Julien K.,  Knobloch E., 1998, JFM, 360, 141
\bibitem[Julien et al.(2012)]{Julien2012} Julien K., Knobloch E., Rubio A. M., Vasil G. M. 2012, PRL, 109, 254503
\bibitem[Julien et al.(2017)]{Julien2017} Julien K., Knobloch E., Plumley M., 2017, JFM, 837, R4
\bibitem[K\"{a}pyl\"{a} et al.(2004)]{Kapyla2004} K\"{a}pyl\"{a} P. J., Korpi M. J., Tuominen I., Chan K. L., 2004, A\& A, 442, 793
\bibitem[K\"{a}pyl\"{a} et al.(2011)]{Kapyla2011} K\"{a}pyl\"{a} P. J., Korpi M. J., Hackman T., 2011, ApJ, 742, 34
\bibitem[K\"{a}pyl\"{a} et al.(2017)]{Kapyla2017} K{\"a}pyl{\"a} P.~J., K{\"a}pyl{\"a} M.~J., Olspert N., Warnecke J., {Brandenburg} A., 2017, A\& A, 599, A4
\bibitem[K\"{a}pyl\"{a} et al.(2017a)]{Kapylaetal2017} K{\"a}pyl{\"a} P.~J., K{\"a}pyl{\"a} M.~J., Rheinhardt M., {Brandenburg} A., Arlt R., K\"{a}pyl\"{a} M. J., Lagg A.,  Olspert N., \& Warnecke, J., 2017, \apjl, 845(2), p.L23
\bibitem[King et al.(2009)]{Kingetal2009} King E. M., Stellmach S., Noir J., Hansen U., Aurnou J. M., 2009, Nature, 457(7227), p.301.
\bibitem[King \& Buffett(2013)]{King2013} King E. M., Buffett B. A., 2013, Earth Planet. Sci. Lett., 371, 156
\bibitem[King et al.(2013)]{KingStellmachBuffett2013} King E.M., Stellmach S., Buffett, B., 2013, Scaling behaviour in Rayleigh‚ÄìB√©nard convection with and without rotation. Journal of Fluid Mechanics, 717, pp.449-471.
\bibitem[Kippenhahn et al.(2012)]{Kip2012} Kippenhahn R., Weigert A.,  Weiss A., 2012, Stellar Structure and Evolution: , Astronomy and Astrophysics Library.~ISBN 978-3-642-30255-8.~Springer-Verlag Berlin Heidelberg, 2012
\bibitem[Korre et al.(2017)]{Korreetal2017} Korre L., Brummell N., Garaud P., 2017., Physical Review E, 96(3), p.033104.
\bibitem[Kraichnan(1962)]{Kraichnan1962} Kraichnan R.~H., 1961, Physics of Fluids, 5(11), 1374
\bibitem[Kupka \& Muthsam(2017)]{KupMuth2017} Kupka F., Muthsam H.~J., 2017, Living Reviews in Computational Astrophysics, 3, 1
\bibitem[Lesaffre et al.(2013)]{Lesaffre2013} Lesaffre P., Chitre S. M., Potter A. T., Tout C. A., 2013, MNRAS, 431, 2200
\bibitem[Malkus(1954)]{Malkus1954} Malkus W.~.V.~R, 1954, Proceedings of the Royal Society of London Series A, 225 (1161), 196
\bibitem[Mantere et al.(2011)]{Mantereetal2011}{Mantere}, M.~J., {K{\"a}pyl{\"a}}, P.~J., {Hackman}, T., 2011, Astronomische Nachrichten, 332, 876
\bibitem[Meakin \& Arnett(2007)]{MeakinArnett2007} Meakin, C.A. and Arnett, D., 2007. Turbulent convection in stellar interiors. I. Hydrodynamic simulation. \apj, 667(1), p.448.
\bibitem[Miesch et al.(2006)]{Mieschetal2006} Miesch M.S., Brun A.S.,  Toomre J., 2006., \apj, 641(1), p.618.
\bibitem[Nordlund et al.(2009)]{Nordlund2009} {Nordlund} {\AA}., {Stein} R.~F., {Asplund} M., 2009, Living Reviews in Solar Physics, 6, 2
\bibitem[Novi et al.(2019)]{Novi2019} {Novi} L., {von Hardenberg} J., {Hughes} D.W., {Provenzale} A., {Spiegel} E.~A., 2019, Phys. Rev. E.,  99, 053116
\bibitem[Raynaud et al.(2018)]{Raynaudetal2018} Raynaud R., Rieutord M., Petitdemange L., Gastine T., Putigny, B., 2018, A\&A, 609, A124
\bibitem[Rempel(2005)]{Rempel2005} Rempel M., 2005, \apj, 622(2), p.1320.
\bibitem[Renzini(1987)]{Renzini1987} Renzini, A., 1987, \aap, 188(1), p.49
\bibitem[Rubio et al.(2014)]{Rubio2014} Rubio A. M., Julien K., Knobloch E., Weiss J. B., 2014, Physical Review Letters, 112, 14, 144501
\bibitem[Schmitz \& Tilgner(2009)]{Tilgner09} Schmitz S., Tilgner A., 2009, Phys. Rev. E, 80, 015305
\bibitem[Spiegel \& Veronis(1960)]{SV1961} Spiegel E.~A., Veronis, G, 1960, ApJ, 131, 442
\bibitem[Spiegel(1971)]{Spiegel1971} Spiegel E.~A., 1971, ARA\& A, 9, 323
\bibitem[Stellmach et al.(2014)]{Stellmach2014} Stellmach S., Lischper M., Julien K., Vasil G., Cheng J. S., Ribeiro A., King E. M., Aurnou J. M., 2014, PRL, 113, 25, id.254501
\bibitem[Stevenson(1979)]{Stevenson1979} Stevenson D. J., 1979, GApFD, 12, 139
\bibitem[Strugarek et al.(2018)]{Strugarek2018} Strugarek A., Beaudoin P., Charbonneau P., Brun A.~S., 2018, ApJ, 863, 35
\bibitem[Teed et al.(2012)]{Teedetal2012} Teed R.J., Jones C.A, Hollerbach R., 2012, Physics of fluids, 24(6), p.066604.
\bibitem[Unno(1967)]{Unno1967} Unno, W., 1967, \pasj 19, p.140
\bibitem[Vallis(2006)]{Vallis2006} Vallis G. K., CUP, 2006
\bibitem[Viallet et al.(2013)]{Vialletetal2013} Viallet, M., Meakin, C., Arnett, D. and Moc√°k, M., 2013. \apj, 769(1), p.1.
\bibitem[Xiong(1978)]{Xiong1978} Xiong, D.~R., 1978, Chin Astron 2:118‚Äì138
\bibitem[Xiong(1989)]{Xiong1989} Xiong, D.R., 1989. Astronomy and Astrophysics, 209, pp.126-134.
\bibitem[Xiong et al.(1997)]{Xiongetal1997} Xiong, D.R., Cheng, Q.L. and Deng, L., 1997. The Astrophysical Journal Supplement Series, 108(2), p.529.
\bibitem[Yadav et al.(2015)]{Yadavetal2015} Yadav R.K., Gastine T., Christensen U.R., Duarte L.D.V., Reiners, A., 2015. Geophysical Journal International, 204(2), pp.1120-1133.


\end{thebibliography}


\appendix

\section{Table of simulations}
\label{AppendixTables}
A summary of the input parameters, and some outputted quantities, for the simulations is given in Table~\ref{Table1}.

\begin{table*}
\caption{A summary of input and output quantities for the simulations described in the main text. Column one gives each simulation a name, for easy reference; the capital letter denotes which setup was used for the simulation. Columns two to six are input quantities, namely the rotation rate, $\Omega$, latitude $\phi$, {\bf horizontal} box sizes $L_x$ and $L_y$ and the kinematic viscosity $\nu$. Column seven gives the resolution of each simulation ($N_x\times N_y \times N_z$) . For simulations with Dedalus (setup A), this is the number of Fourier (Chebyshev) modes in the horizontal (vertical) directions after de-aliasing has been applied (i.e., the total number of modes in each direction is 3/2 times the numbers quoted here). For the Nek5000 simulations (setup B), we give the total number of points in each dimension using a polynomial order 10 in each element (using 3/2 times the number quoted here for the nonlinear terms). 
The data in the final five columns are derived from the simulation results, more traditional non-dimensional numbers (such as those given in (\ref{tradnondim})) can be obtained from quantities given in the table.
The definitions of $|\frac{d\langle T\rangle }{dz}|$, $u_{z,rms}$, and $\delta T_{rms}$ are given in \S\ref{dTdzsims} and \S\ref{subsec:compOQ} respectively. 
The Rossby number $Ro$ is defined by $Ro=u_{z,rms}/(2\Omega H)$,
and the vertical Reynolds number, $Re_z$ from $Re_z=u_{z,rms}H/\nu$.
Our simulation units are determined by setting $F=H=1$. To restore units one should replace $\Omega\rightarrow \Omega H^{2/3}/F^{1/3}$, $L_\perp\rightarrow L_\perp/H$,$\nu\rightarrow \nu/(F^{1/3}H^{4/3})$, $v_z\rightarrow v_z/(FH)^{1/3}$, and ${dT/ dz}\rightarrow {dT/dz}H^{4/3}/F^{2/3}$. Note that $\nu=\kappa$ is assumed throughout.
}\label{Table1}

\begin{tabular}{ccccccc||ccccc}

\hline
Name &$\Omega$ & $\phi$ & $L_x$ & $L_y$ & $-\log_{10}{\nu}$ & Resolution & $|\frac{d\langle T\rangle }{dz}|$ & $u_{z,rms}$ & $\delta T_{rms}$ & $Ro$ & $Re_z$ 
 \\
 \hline
\hline 
6B80a & 6 & $80^{\circ}$ & 2 & 2 & 2.9 & $200 \times 200 \times 200$ & $8.0\pm 0.5$ & 0.64$\pm 0.03$ & 2.16$\pm0.12$ & $5.33 \times 10^{-2}$ & 508 \\ 
6B70a & 6 & $70^{\circ}$ & 2 & 2 & 2.9 & $200 \times 200 \times 200$ & $7.93\pm 0.63$ & 0.65$\pm 0.04$ & 2.21$\pm0.14$ & $5.41 \times 10^{-2}$ & 516 \\ 
6B60a & 6 & $60^{\circ}$ & 2 & 2 & 2.9 & $200 \times 200 \times 200$ & $8.39\pm 0.61$ & 0.64$\pm 0.04$ & 2.29$\pm0.14$ & $5.33 \times 10^{-2}$  & 508 \\ 
6B50a & 6 & $50^{\circ}$ & 2 & 2 & 2.9 & $200 \times 200 \times 200$ & $8.3\pm 0.77$ & 0.64$\pm 0.04$ & 2.40$\pm0.16$ & $5.33 \times 10^{-2}$  & 508 \\ 
6B40a & 6 & $40^{\circ}$ & 2 & 2 & 2.9 & $200 \times 200 \times 200$ & $8.3\pm 1.1$ & 0.64$\pm 0.06$ & 2.50$\pm0.28$ & $5.33 \times 10^{-2}$  & 508 \\ 
6B30a & 6 & $30^{\circ}$ & 2 & 2 & 2.9 & $200 \times 200 \times 200$ & $6.7\pm 1.62$ & 0.7$\pm 0.1$ & 2.52$\pm0.42$ & $5.83 \times 10^{-2}$  & 556 \\ 
6B30b & 6 & $30^{\circ}$ & 3 & 3 & 2.9 & $200 \times 200 \times 200$ & $7.17\pm 0.92$ & 0.67$\pm 0.06$ & 2.58$\pm0.29$ & $5.58 \times 10^{-2}$ & 532 \\ 
6B20a & 6 & $20^{\circ}$ & 2 & 2 & 2.9 & $200 \times 200 \times 200$ & $3.05\pm 1.77$ & 0.91$\pm 0.15$ & 2.09$\pm0.43$ & $7.58 \times 10^{-2}$  & 723 \\ 
\hline
10B90a & 10 & $90^{\circ}$ & 1 & 1 & 2.9 & $200 \times 200 \times 200$ & $10.8\pm 0.98$ & 0.59$\pm 0.06$ & 2.44$\pm0.18$ & $2.95 \times 10^{-2}$ & 469 \\ 
10B85a & 10 & $85^{\circ}$ & 2 & 1 & 2.9 & $200 \times 200 \times 200$ & $11.26\pm 0.65$ & 0.58$\pm 0.04$ & 2.29$\pm0.17$ & $2.90 \times 10^{-2}$  & 461 \\ 
10B75a & 10 & $75^{\circ}$ & 1 & 1 & 2.9 & $200 \times 200 \times 200$ & $10.8\pm 0.98$ & 0.61$\pm 0.06$ & 2.44$\pm0.18$ & $3.05 \times 10^{-2}$ & 485 \\
10B75b & 10 & $75^{\circ}$ & 2 & 1 & 2.9 & $200 \times 200 \times 200$ & $11.31\pm 0.76$ & 0.58$\pm 0.04$ & 2.44$\pm0.17$ & $2.90 \times 10^{-2}$ & 461 \\ 
10B75c & 10 & $75^{\circ}$ & 2 & 2 & 2.9 & $200 \times 200 \times 200$ & $11.34\pm 0.69$ & 0.58$\pm 0.03$ & 2.42$\pm0.14$ & $2.90 \times 10^{-2}$ &  461 \\
10B75d & 10 & $75^{\circ}$ & 4 & 2 & 2.9 & $200 \times 200 \times 200$ & $11.26\pm 0.35$ & 0.59$\pm 0.02$ & 2.45$\pm0.07$ & $2.95 \times 10^{-2}$ & 469 \\ 
10B70a & 10 & $70^{\circ}$ & 2 & 2 & 2.9 & $200 \times 200 \times 200$ & $11.12\pm 0.57$ & 0.60$\pm 0.03$ & 2.44$\pm0.13$ & $3.00 \times 10^{-2}$ & 477 \\ 
10B60a & 10 & $60^{\circ}$ & 1 & 1 & 2.9 & $200 \times 200 \times 200$ & $11.6\pm 1.12$ & 0.60$\pm 0.06$ & 2.59$\pm0.22$ & $3.00 \times 10^{-2}$ &  477 \\
10B60b & 10 & $60^{\circ}$ & 1 & 1.5 & 2.9 & $200 \times 200 \times 200$ & $11.3\pm 0.89$ & 0.58$\pm 0.05$ & 2.55$\pm0.23$ & $2.90 \times 10^{-2}$ & 461 \\
10B60c & 10 & $60^{\circ}$ & 1 & 2 & 2.9 & $200 \times 200 \times 200$ & $11.86\pm 0.82$ & 0.57$\pm 0.04$ & 2.59$\pm0.19$ & $2.85 \times 10^{-2}$ & 453 \\
10B60d & 10 & $60^{\circ}$ & 2 & 2 & 2.9 & $200 \times 200 \times 200$ & $11.46\pm 0.61$ & 0.59$\pm 0.03$ & 2.53$\pm0.14$ & $2.95 \times 10^{-2}$ &  469 \\
10B60e & 10 & $60^{\circ}$ & 4 & 4 & 2.9 & $200 \times 200 \times 200$ & $11.41\pm 0.25$ & 0.59$\pm 0.01$ & 2.58$\pm0.07$ & $2.95 \times 10^{-2}$ &  469 \\
10B55a & 10 & $55^{\circ}$ & 2 & 2 & 2.9 & $200 \times 200 \times 200$ & $11.23\pm 0.59$ & 0.59$\pm 0.03$ & 2.60$\pm0.15$ & $2.95 \times 10^{-2}$ &  469 \\ 
10B45a & 10 & $45^{\circ}$ & 1 & 1 & 2.9 & $200 \times 200 \times 200$ & $10.59\pm 0.25$ & 0.60$\pm 0.08$ & 2.74$\pm0.37$ & $3.00 \times 10^{-2}$ &  477 \\
10B45b & 10 & $45^{\circ}$ & 1.5 & 1.5 & 2.9 & $200 \times 200 \times 200$ & $11.9\pm 1.14$ & 0.57$\pm 0.05$ & 2.81$\pm0.33$ & $2.85 \times 10^{-2}$ &  453 \\
10B45c & 10 & $45^{\circ}$ & 1 & 2 & 2.9 & $200 \times 200 \times 200$ & $12.42\pm 1.24$ & 0.55$\pm 0.06$ & 2.83$\pm0.3$ & $2.75 \times 10^{-2}$ &  437 \\
10B45d & 10 & $45^{\circ}$ & 1 & 4 & 2.9 & $200 \times 200 \times 200$ & $12.7\pm 0.83$ & 0.55$\pm 0.04$ & 2.86$\pm0.2$ & $2.75 \times 10^{-2}$ & 437 \\
10B45e & 10 & $45^{\circ}$ & 2 & 2 & 2.9 & $200 \times 200 \times 200$ & $11.86\pm 0.75$ & 0.57$\pm 0.04$ & 2.78$\pm0.2$ & $2.85 \times 10^{-2}$ & 453 \\
10B45f & 10 & $45^{\circ}$ & 4 & 2 & 2.9 & $200 \times 200 \times 200$ & $10.99\pm 0.57$ & 0.58$\pm 0.03$ & 2.75$\pm0.14$ & $2.90 \times 10^{-2}$ & 461 \\
10B45g & 10 & $45^{\circ}$ & 4 & 4 & 2.9 & $200 \times 200 \times 200$ & $11.36\pm 0.34$ & 0.58$\pm 0.02$ & 2.77$\pm0.11$ & $2.90 \times 10^{-2}$ &  461 \\
10B45h & 10 & $45^{\circ}$ & 10 & 10 & 2.9 & $400 \times 400 \times 200$ & $11.1\pm 0.14$ & 0.58$\pm 0.01$ & 2.77 & $2.90 \times 10^{-2}$ &  461 \\
10B40a & 10 & $40^{\circ}$ & 2 & 2 & 2.9 & $200 \times 200 \times 200$ & $11.36\pm 0.69$ & 0.58$\pm 0.04$ & 2.67$\pm0.15$ & $2.90 \times 10^{-2}$ &  461 \\ 
10B30a & 10 & $30^{\circ}$ & 2 & 2 & 2.9 & $200 \times 200 \times 200$ & $9.8\pm 1.8$ & 0.55$\pm 0.08$ & 2.98$\pm0.14$ & $2.75 \times 10^{-2}$ &  437 \\
10B30b & 10 & $30^{\circ}$ & 1 & 4 & 2.9 & $200 \times 200 \times 200$ & $15.4\pm 1.5$ & 0.48$\pm 0.05$ & 3.37$\pm0.34$ & $2.40 \times 10^{-2}$ &  381 \\
10B30c & 10 & $30^{\circ}$ & 8 & 8 & 2.9 & $200 \times 200 \times 200$ & $9.08\pm 0.35$ & 0.62$\pm 0.02$ & 2.91$\pm0.08$ & $3.10 \times 10^{-2}$ &  492 \\
10B30d & 10 & $30^{\circ}$ & 10 & 10 & 2.9 & $400 \times 400 \times 200$ & $8.82\pm 0.37$ & 0.63$\pm 0.02$ & 2.92$\pm0.08$ & $3.15 \times 10^{-2}$ &  500 \\
10B15a & 10 & $15^{\circ}$ & 10 & 10 & 2.9 & $400 \times 400 \times 200$ & $7.04\pm 1.58$ & 0.66$\pm 0.07$ & 2.99$\pm0.21$ & $3.30 \times 10^{-2}$ &  524 \\
10B10a & 10 & $10^{\circ}$ & 10 & 10 & 2.9 & $400 \times 400 \times 200$ & $1.6\pm 3.0$ & 1.04$\pm 0.24$ &  & $5.20 \times 10^{-2}$ & 826 \\
\hline

10A90a & 10 & $90^{\circ}$ & 1 & 1 & 3.3     & $128 \times 128 \times 256$  & $12.95\pm 1.13$ & $0.60\pm 0.06$& $2.47\pm 0.22$ & $3.02 \times 10^{-2}$ &  1205\\
10A90b & 10 & $90^{\circ}$ & 1 & 1 & 3       & $128 \times 128 \times 128$  & $12.17\pm 0.98$ & $0.57\pm 0.05$& $2.43\pm 0.20$ & $2.86 \times 10^{-2}$ &  571 \\
10A90c & 10 & $90^{\circ}$ & 2 & 2 & 3       & $128 \times 128 \times 128$  & $12.83\pm0.66$ & $0.57\pm0.03$& $2.96\pm0.13$ & $2.86 \times 10^{-2}$ & 572 \\
10A75a & 10 & $75^{\circ}$ & 1 & 1 & 3.3     & $128 \times 128 \times 256$ & $13.43\pm 1.01$ & $0.60\pm 0.05$& $2.53\pm 0.23$ & $3.01 \times 10^{-2}$ &  1199 \\
10A75b & 10 & $75^{\circ}$ & 1   & 2   & 3   & $128 \times 128 \times 128$ & $13.16\pm0.90$ & $0.56\pm0.03$& $2.56\pm0.16$ & $2.82 \times 10^{-2}$ & 563 \\
10A75c & 10 & $75^{\circ}$ & 2   & 1   & 3   & $128 \times 128 \times 128$ & $12.87\pm0.76$ & $0.57\pm0.03$& $2.53\pm0.16$ & $2.84 \times 10^{-2}$&  568 \\
10A75d & 10 & $75^{\circ}$ & 2   & 2   & 3   & $128 \times 128 \times 128$ & $12.57\pm0.63$ & $0.59\pm0.03$& $2.70\pm0.12$ & $2.96 \times 10^{-2}$ &  592 \\
10A60a & 10 & $60^{\circ}$ & 1   & 1   & 3.3 & $128 \times 128 \times 256$ & $13.73\pm 1.21$ & $0.60\pm 0.05$& $2.68\pm 0.24$ & $3.00 \times 10^{-2}$ & 1197 \\
10A60b & 10 & $60^{\circ}$ & 1   & 2   & 3   & $128 \times 128 \times 128$ & $13.57\pm1.01$ & $0.55\pm0.04$& $2.69\pm0.18$ & $2.77 \times 10^{-2}$ &  554 \\
10A60c & 10 & $60^{\circ}$ & 2   & 1   & 3   & $128 \times 128 \times 128$ & $13.01\pm0.92$ & $0.57\pm0.03$& $2.63\pm0.17$ & $2.83 \times 10^{-2}$ &  566 \\
10A60d & 10 & $60^{\circ}$ & 2   & 2   & 3   & $128 \times 128 \times 128$ & $12.74\pm0.70$ & $0.58\pm0.03$& $2.62\pm0.14$ & $2.91 \times 10^{-2}$ &  582 \\
10A60e & 10 & $60^{\circ}$ & 1.5 & 1.5 & 3   & $128 \times 128 \times 128$ & $12.94\pm0.85$ & $0.57\pm0.03$& $2.62\pm0.15$ & $2.83 \times 10^{-2}$ &  565\\
10A45a & 10 & $45^{\circ}$ & 1   & 1   & 3.3 & $128 \times 128 \times 256$ & $13.22\pm 1.54$ & $0.59\pm 0.06$& $2.87\pm 0.34$ & $2.97 \times 10^{-2}$ &1185 \\
10A45b & 10 & $45^{\circ}$ & 1.5 & 1.5 & 3   & $128 \times 128 \times 128$ & $13.13\pm1.21$ & $0.56\pm0.04$& $2.84\pm0.21$ & $2.79 \times 10^{-2}$ &  562 \\
10A45c & 10 & $45^{\circ}$ & 2   & 2   & 3   & $128 \times 128 \times 128$ & $12.96\pm0.94$ & $0.56\pm0.03$& $2.84\pm0.16$ & $2.81 \times 10^{-2}$ & 558 \\
10A45d & 10 & $45^{\circ}$ & 2   & 2   & 3   & $128 \times 128 \times 128$ & $13.21\pm0.94$ & $0.56\pm0.04$& $2.85\pm0.18$ & $2.79 \times 10^{-2}$ & 558  \\

\end{tabular}

\end{table*}

\begin{table*}
\contcaption{}
\begin{tabular}{ccccccc||ccccc}
\hline
Name &$\Omega$ & $\phi$ & $L_x$ & $L_y$ & $-\log_{10}{\nu}$ & Resolution & $|\frac{d\langle T\rangle }{dz}|$ & $u_{z,rms}$ & $\delta T_{rms}$ & $Ro$ & $Re_z$ 
 \\
 \hline
\hline

10A45e & 10 & $45^{\circ}$ & 2   & 4   & 3   & $128 \times 384 \times 128$ & $15.22\pm0.68$ & $0.52\pm0.02$& $3.10\pm0.13$ & $2.62 \times 10^{-2}$ & 525 \\
10A45f & 10 & $45^{\circ}$ & 2   & 6   & 3   & $192 \times 384 \times 128$ & $13.96\pm0.56$ & $0.54\pm0.02$& $2.93\pm0.09$ & $2.68 \times 10^{-2}$ &  535 \\
10A45g & 10 & $45^{\circ}$ & 2   & 8   & 3   & $128 \times 768 \times 128$ & $14.74\pm0.42$ & $0.53\pm0.02$& $2.96\pm0.08$ & $2.67 \times 10^{-2}$ &  534 \\
10A45h & 10 & $45^{\circ}$ & 4   & 2   & 3   & $384 \times 192 \times 128$ & $12.75\pm0.61$ & $0.56\pm0.02$& $2.84\pm0.12$ & $2.81 \times 10^{-2}$ &  563 \\
10A45i & 10 & $45^{\circ}$ & 4   & 4   & 3   & $384 \times 384 \times 128$ & $12.90\pm0.47$ & $0.57\pm0.02$& $2.84\pm0.08$ & $2.83 \times 10^{-2}$ &  566 \\
10A45k & 10 & $45^{\circ}$ & 8   & 2   & 3   & $768 \times 192 \times 128$ & $12.67\pm0.43$ & $0.57\pm0.01$& $2.86\pm0.09$ & $2.83 \times 10^{-2}$ &  566 \\
10A30a & 10 & $30^{\circ}$ & 1   & 1   & 3.3 & $128 \times 128 \times 256$ & $9.46\pm2.71$  & $0.69\pm0.13$& $2.79\pm0.54$ & $3.45 \times 10^{-2}$ &  1377 \\
10A30b & 10 & $30^{\circ}$ & 2   & 2   & 2.5 & $128 \times 128 \times 128$ & $10.73\pm1.92$ & $0.51\pm0.06$& $3.00\pm0.34$ & $2.54 \times 10^{-2}$ &  161 \\
10A30c & 10 & $30^{\circ}$ & 2.5 & 2.5 & 2.5 & $128 \times 128 \times 128$ & $11.85\pm1.30$ & $0.49\pm0.05$& $3.10\pm0.28$ & $2.46 \times 10^{-2}$ & 155  \\
10A30d & 10 & $30^{\circ}$ & 1   & 3   & 3   & $128 \times 256 \times 128$ & $17.12\pm1.64$ & $0.49\pm0.05$& $3.42\pm0.36$ & $2.47 \times 10^{-2}$ &  493 \\
10A30e & 10 & $30^{\circ}$ & 3   & 1   & 3   & $256 \times 128 \times 128$ & $9.17\pm 1.69$ & $0.64\pm0.08$& $2.85\pm0.40$ & $3.18 \times 10^{-2}$ &  636 \\
10A20a & 10 & $20^{\circ}$ & 4 & 4 & 3 & $256 \times 256 \times 128$ & $12.90\pm1.56$ & $0.59\pm0.06$& $3.45\pm0.24$  & $2.95 \times 10^{-2}$ & 589 \\
10A15a & 10 & $15^{\circ}$ & 5 & 5 & 3 & $128 \times 384 \times 384$ & $6.21\pm1.26$  & $0.73\pm0.07$& $3.01\pm0.24$  &  $3.67 \times 10^{-2}$ & 734 \\

\hline 
20B80a & 20 & $80^{\circ}$ & 1.5 & 1.5 & 3.3 & $200 \times 200 \times 200$ & $18.9\pm 0.52$ & 0.59$\pm 0.03$ & 2.80$\pm0.14$ & $1.48 \times 10^{-2}$ & 1177 \\
20B70a & 20 & $70^{\circ}$ & 1.5 & 1.5 & 3.3 & $200 \times 200 \times 200$ & $18.2\pm 0.49$ & 0.61$\pm 0.03$ & 2.84$\pm0.13$ & $1.53 \times 10^{-2}$ &  1217 \\
20B60a & 20 & $60^{\circ}$ & 1.5 & 1.5 & 3.3 & $200 \times 200 \times 200$ & $18.7\pm 0.53$ & 0.58$\pm 0.03$ & 2.96$\pm0.17$ & $1.45 \times 10^{-2}$ & 1157 \\
20B50a & 20 & $50^{\circ}$ & 1.5 & 1.5 & 3.3 & $200 \times 200 \times 200$ & $18.8\pm 0.74$ & 0.55$\pm 0.03$ & 3.1$\pm0.19$ & $1.38 \times 10^{-2}$ & 1097\\
20B40a & 20 & $40^{\circ}$ & 1.5 & 1.5 & 3.3 & $200 \times 200 \times 200$ & $21.8\pm 1.0$ & 0.51$\pm 0.04$ & 3.48$\pm0.26$ & $1.28 \times 10^{-2}$ &  1018 \\
20B40b & 20 & $40^{\circ}$ & 3.0 & 3.0 & 3.3 & $400 \times 400 \times 200$ & $18.13\pm 1.2$ & 0.52$\pm 0.02$ & 3.27$\pm0.15$ & $1.30 \times 10^{-2}$ &1038   \\
20B30a & 20 & $30^{\circ}$ & 1.5 & 1.5 & 3.3 & $200 \times 200 \times 200$ & $22.2\pm 1.47$ & 0.56$\pm 0.01$ & 3.79$\pm0.44$ & $1.40 \times 10^{-2}$ & 1117 \\
20B30b & 20 & $30^{\circ}$ & 3.0 & 3.0 & 3.3 & $400 \times 400 \times 200$ & $20.4\pm 1.72$ & 0.48$\pm 0.03$ & 3.65$\pm0.15$ & $1.20 \times 10^{-2}$ &  958 \\
20B20a & 20 & $20^{\circ}$ & 1.5 & 1.5 & 3.3 & $200 \times 200 \times 200$ & $8.38\pm 2.0$ & 0.69$\pm 0.13$ & 2.99$\pm0.61$ & $1.73 \times 10^{-2}$  & 1377 \\
\hline
30B75a & 30 & $75^{\circ}$ & 0.6 & 0.6 & 3.8 & $200 \times 200 \times 200$ & $27.7\pm 1.47$ & 0.56$\pm 0.08$ & 2.94$\pm0.29$ & $9.33 \times 10^{-3}$ & 3533  \\
30B60a & 30 & $60^{\circ}$ & 0.6 & 0.6 & 3.8 & $200 \times 200 \times 200$ & $28.2\pm 2.05$ & 0.53$\pm 0.07$ & 3.20$\pm0.31$ & $8.83 \times 10^{-3}$ & 3344\\
30B60b & 30 & $60^{\circ}$ & 1 & 1 & 3.3 & $200 \times 200 \times 200$ & $25.3\pm 1.05$ & 0.49$\pm 0.04$ & 3.31 & $8.17 \times 10^{-3}$ &  978 \\
30B60c & 30 & $60^{\circ}$ & 1 & 1.5 & 3.3 & $200 \times 200 \times 200$ & $27.6\pm 0.8$ & 0.48$\pm 0.03$ & 3.36 & $8.00 \times 10^{-3}$ & 958 \\
30B45a & 30 & $45^{\circ}$ & 0.6 & 0.6 & 3.2& $200 \times 200 \times 200$ & $25.7\pm 1.82$ & 0.54$\pm 0.07$ & 3.82 & $9.00 \times 10^{-3}$ & 856  \\
30B45b & 30 & $45^{\circ}$ & 0.6 & 1 & 3.3 & $200 \times 200 \times 200$ & $26.0\pm 1.45$ & 0.48$\pm 0.05$ & 3.57$\pm0.41$ & $8.00 \times 10^{-3}$ & 958 \\
30B45c & 30 & $45^{\circ}$ & 0.6 & 1.5 & 3.3 & $200 \times 200 \times 200$ & $29.1\pm 1.12$ & 0.45$\pm 0.04$ & 3.86$\pm0.26$ & $7.50 \times 10^{-3}$ & 898 \\
30B45d & 30 & $45^{\circ}$ & 0.6 & 1.2 & 3.3 & $200 \times 200 \times 200$ & $28.15\pm 1.2$ & 0.45$\pm 0.04$ & 3.71$\pm0.38$ & $7.50 \times 10^{-3}$ & 898 \\
30B30a & 30 & $30^{\circ}$ & 0.8 & 0.8 & 3.2 & $200 \times 200 \times 200$ & $18.6\pm 3.6$ & 0.55$\pm 0.13$ & 3.32$\pm0.82$ & $9.17 \times 10^{-3}$ & 872\\
30B30b & 30 & $30^{\circ}$ & 0.8 & 0.8 & 3.3 & $200 \times 200 \times 200$ & $16.6\pm 2.9$ & 0.54$\pm 0.14$ & 3.32$\pm0.82$ & $9.00 \times 10^{-3}$ & 1077  \\
30B20a & 30 & $20^{\circ}$ & 0.8 & 0.8 & 3.3 & $200 \times 200 \times 200$ & $6.6\pm 2.4$ & 0.65$\pm 0.17$ & 2.55$\pm0.70$ & $1.08 \times 10^{-2}$  & 1297  \\

\hline 
30A90a & 30 & $90^{\circ}$   & 0.4 & 0.4 & 3.5 & $128 \times 128 \times 128$ & $26.50\pm1.89$ & $0.52\pm0.07$&  & $8.64 \times 10^{-3}$ & 1640 \\
30A90b & 30 & $90^{\circ}$   & 0.8 & 0.8 & 3.2 & $128 \times 128 \times 128$ & $27.77\pm0.89$ & $0.48\pm0.03$& $3.10\pm0.21$ & $8.06 \times 10^{-3}$ & 766 \\
30A90c & 30 & $90^{\circ}$   & 1.2 & 1.2 & 3.2 & $192 \times 192 \times 192$ & $28.30\pm0.62$ & $0.49\pm0.02$& $3.23\pm0.17$ & $8.09 \times 10^{-3}$ & 769 \\
30A75a & 30 & $75^{\circ}$   & 0.4 & 0.4 & 3.8 & $128 \times 128 \times 128$ & $27.15\pm1.59$ & $0.55\pm0.07$& $3.03\pm0.39$ & $9.15 \times 10^{-3}$& 3462 \\
30A75b & 30 & $75^{\circ}$   & 0.8 & 0.8 & 3.5 & $192 \times 192 \times 192$ & $28.24\pm1.13$ & $0.55\pm0.04$& $3.16\pm0.19$ & $9.13 \times 10^{-3}$&  1733 \\

30A60a & 30 & $60^{\circ}$   & 0.4 & 0.4 & 3.8 & $128 \times 128 \times 128$ & $28.59\pm1.95$ & $0.53\pm0.06$& $3.35\pm0.42$ &$8.80 \times 10^{-3}$ &  3333 \\
30A60b & 30 & $60^{\circ}$   & 0.8 & 0.8 & 3.5 & $192 \times 192 \times 192$ & $28.74\pm1.08$ & $0.52\pm0.05$& $3.36\pm0.28$ &$8.63 \times 10^{-3}$ &  1637 \\
30A45a & 30 & $45^{\circ}$  & 0.6 & 0.6 & 3.5 & $128 \times 128 \times 128$ & $27.94\pm1.78$ & $0.52\pm0.07$& $3.65\pm0.44$ & $8.75 \times 10^{-3}$ & 1660 \\
30A45b & 30 & $45^{\circ}$  & 1.5 & 1.5 & 3.5 & $192 \times 192 \times 192$ & $31.13\pm0.99$ & $0.45\pm0.02$& $3.80\pm0.19$ & $7.55 \times 10^{-3}$ & 904 \\

30A30a & 30 & $30^{\circ}$   & 0.8 & 0.8 & 3.2 & $192 \times 192 \times 192$ &$20.06\pm3.69$ & $0.55\pm0.15$ & $3.77\pm0.98$ & $9.20 \times 10^{-3}$  & 875 \\
30A30b & 30 & $30^{\circ}$   & 2   & 2   & 3.2 & $256 \times 256 \times 192$ & $33.48\pm1.48$ & $0.42\pm0.05$& $4.32\pm0.45$ & $7.02 \times 10^{-3}$  & 667 \\

30A26a & 30 & $25.7^{\circ}$ & 0.8 & 0.8 & 3.3 & $128 \times 128 \times 128$ & $16.12\pm2.78$ & $0.56\pm0.12$& $3.55\pm0.83$ & $9.36 \times 10^{-3}$ & 1120 \\
30A26b & 30 & $25.7^{\circ}$ & 1.6 & 1.6 & 3.3 & $128 \times 128 \times 128$ & $29.69\pm3.13$ & $0.50\pm0.09$& $4.22\pm0.51$ & $8.28 \times 10^{-3}$ & 991 \\

\hline
50B75a & 50 & $75^{\circ}$  & 1 & 1 & 3.6 & $400 \times 400 \times 200$ & $37.3\pm0.85$ & $0.51\pm0.03$  & $3.38\pm0.16$  & $5.10 \times 10^{-3}$ & 2030  \\
50B60a & 50 & $60^{\circ}$  & 1 & 1 & 3.6 & $400 \times 400 \times 200$ & $36.5\pm0.66$ & $0.50\pm0.02$  & $3.51\pm0.17$  & $5.00 \times 10^{-3}$ & 1991  \\
50B45a & 50 & $45^{\circ}$  & 1 & 1 & 3.6 & $400 \times 400 \times 200$ & $39.0\pm0.8$ & $0.45\pm0.05$  & $3.98\pm0.27$  & $4.50 \times 10^{-3}$ & 1791  \\
50B30a & 50 & $30^{\circ}$  & 0.6 & 0.6 & 3.6 & $200 \times 200 \times 200$ & $26.1\pm4.9$ & $0.52\pm0.16$  &   & $5.20 \times 10^{-3}$ & 2070  \\

\hline
70B75a & 70 & $75^{\circ}$  & 1 & 1 & 3.6 & $400 \times 400 \times 200$ & $49.0\pm0.37$ & $0.44\pm0.02$  & $3.57\pm0.13$  & $3.14 \times 10^{-3}$ &  1752 \\
70B45a & 70 & $45^{\circ}$  & 1 & 1 & 3.6 & $400 \times 400 \times 200$ & $43.3\pm0.75$ & $0.43\pm0.03$  & $3.96\pm0.28$  & $3.07 \times 10^{-3}$ &  1712 \\

\hline
100B75a & 100 & $75^{\circ}$  & 0.5 & 0.5 & 4 & $300 \times 300 \times 400$ & $67.4\pm1.3$ & $0.44\pm0.04$  & $3.97\pm0.22$  & $2.20 \times 10^{-3}$  &  4400 \\
100B60a & 100 & $60^{\circ}$  & 0.5 & 0.5 & 4 & $300 \times 300 \times 400$ & $63.1\pm0.9$ & $0.45\pm0.04$  & $4.09\pm0.24$  & $2.25 \times 10^{-3}$ & 4500  \\
100B45a & 100 & $45^{\circ}$  & 0.3 & 0.3 & 4 & $200 \times 200 \times 200$ & $64.2\pm2.2$ & $0.43\pm0.01$  &   &  $2.15 \times 10^{-3}$ & 4300  \\
100B30a & 100 & $30^{\circ}$  & 0.4 & 0.4 & 4 & $200 \times 200 \times 200$ & $45.8\pm5.3$ & $0.46\pm0.12$  &   &  $2.30 \times 10^{-3}$ &  4600 \\
\hline
\hline
\end{tabular}

\end{table*}

\section{Single-mode theory}\label{AppendixSingleMode}
Here we derive the scaling predictions of RMLT as determined by S79 by assuming that a single mode dominates the transport. S79 approximates the total flux in equation~(\ref{Flux}) by considering only the single mode that maximises $F_\lambda$. This produces simple analytical predictions for the bulk properties, which can be obtained by maximising $F_\lambda$ over each of $k_x$, $k_y$ and $n$. Following this procedure, we obtain the corresponding maximal flux $F_{max}$, the linear growth rate $\sigma_{max}$, the wavenumbers ($k_{x,max}, k_{y,max},n\pi$) and the RMS vertical velocity $v_{z,max}$:
\begin{eqnarray}
\label{S791}
&& F_{max} = C_1\frac{N^3_*}{1+4\Omega_z^2/N^2_*}, \;\;\;\;
\label{S792}
\sigma_{max} =\frac{3}{5}N_*, \;\;\;\;\; \\
\label{S793}
&& k^2_{x,max}=\frac{3\pi^2}{2}\left(1+\frac{20}{3}\frac{\Omega_z^2}{N^2_*}\right), \;\;\;\;\;
\label{S794}
 k_{y,max}=0, \;\;\;\;\; \\
\label{S795}
&& n_{max}=1, \;\;\;\;\;
v_{z,max} = C_2 \frac{N_*}{\sqrt{\pi^2+k_{x,max}^2}}.
\label{S796}
\end{eqnarray}
The numerical coefficients here agree with those in S79 Eq.~40.
These predict the mode that transports the most heat in rotating convection, and contain two constants $C_1$ and $C_2$. The modes that transport the most heat are $y$-aligned rolls that do not vary along the horizontal component of $\boldsymbol{\Omega}$. As a result, the predictions depend only on $\Omega_z$, and not on $\Omega_y$.

We can manipulate Eqs.~\ref{S791}--\ref{S796} to determine the dependence of the bulk properties in rotating convection on $\Omega_z$ for a given imposed heat flux $F$. To do so, we assume that this single mode dominates, so that $F_{max}=F$, and solve Eqs.~\ref{S791}--\ref{S796} to determine $N_*^2$ and $v_{z}$, then use Eq.~\ref{MLTheat} to determine the temperature fluctuation $\delta T$. To simplify Eqs.\ref{S791}--\ref{S796} and for comparison with paper I, we consider the limit of rapid rotation ($\Omega_z^2/N^2\gg 1$) and re-introduce the vertical length-scale $H$. This allows us to obtain simple scaling relations describing how the mean temperature gradient, vertical velocity and temperature fluctuations, and the horizontal wavenumber of the modes that dominate heat transport, scale with $\Omega_z$, heat flux $F$ and the depth of convective zone $H$:
\begin{eqnarray}
\label{rotMLT1}
&& 
N_*^2 \sim \frac{F^{\frac{2}{5}}\Omega_z^{\frac{4}{5}}}{H^{\frac{4}{5}}}, \;\;\;\;\;
\label{rotMLT2}
v_z \sim \frac{H^{\frac{1}{5}}F^{\frac{2}{5}}}{\Omega_z^{\frac{1}{5}}},
\label{rotMLT3} \\
&& \delta T \sim \frac{\Omega_z^{\frac{1}{5}}F^{\frac{3}{5}}}{H^{\frac{1}{5}}}, \;\;\;\;\;
\label{rotMLT4}
k_\perp \sim k_{y} \sim \frac{\Omega_z^{\frac{3}{5}}}{F^{\frac{1}{5}}H^{\frac{3}{5}}}.
\end{eqnarray}
These relations were obtained in a different but equivalent way at the poles (where $\Omega_z=\Omega$) using simple physical arguments in Paper I. In particular, the key points are that the convection is dominated by the mode that transports the most heat, that a mode grows for a cascade time (Eq.~\ref{amp}), and that as hot plumes rise, they carry the background temperature gradient for a time $1/\sigma~\sim 1/N_*$ before cascading (Eq.~\ref{fluxdef2}).

Excellent agreement was obtained between this simple single-mode theory and numerical simulations of rapidly rotating convection at the poles in Paper I, over more than two orders of magnitude in rotation rate for cases in which rotation and gravity were aligned\footnote{These results have also been confirmed using setup A, but to save space we have not presented these results.}. The single-mode RMLT presented here suggests that when the direction of gravity is oblique to the rotation vector, i.e., at non-polar latitudes, we should simply vary $\Omega_z$ in these relations. However, we demonstrate in section \ref{compsim} that the single-mode theory above works only approximately, and there is a systematic departure from this theory which we have accounted for by including all of the modes in a multi-mode theory.


\bsp	
\label{lastpage}
\end{document}